\documentclass[journal,article,submit,pdftex,moreauthors]{mdpi}
\usepackage{amsmath,amsfonts}

\usepackage{algorithm}
\usepackage{algorithmic}

\usepackage{subfig}

\usepackage{graphicx}
\firstpage{1} 
\pubvolume{1}
\issuenum{1}
\articlenumber{0}
\pubyear{2024}
\copyrightyear{2024}
\datereceived{ } 
\daterevised{ } % Comment out if no revised date
\dateaccepted{ } 
\datepublished{ } 
\hreflink{https://doi.org/} % If needed use \linebreak
\Title{DTPPO: Dual-Transformer Encoder-based Proximal Policy Optimization for Multi-UAV Navigation in Unseen Complex Environments}

\TitleCitation{Title}

\Author{Anning Wei $^{1, \ddagger}$, Jintao Liang $^{2,\ddagger}$, Kaiyuan Lin $^{3}$, Ziyue Li $^{4}$*,  Rui Zhao $^{5}$}

\AuthorNames{Firstname Lastname, Firstname Lastname and Firstname Lastname}

\AuthorCitation{Lastname, F.; Lastname, F.; Lastname, F.}
\address{%
$^{1}$ \quad Department of Automation, Tsinghua University; weian23@mails.tsinghua.edu.cn and qhdai@tsinghua.edu.cn\\
$^{2}$ \quad Beijing University of Posts and Telecommunications; ljt2021@bupt.edu.cn\\
$^{3}$ \quad University of Southern California; linkaiyu@usc.edu\\
$^{4}$ \quad University of Cologne; zlibn@wiso.uni-koeln.de\\
$^{5}$ \quad SenseTime Research; zhaorui@sensetime.com
}

\corres{Correspondence: zlibn@wiso.uni-koeln.de}

\secondnote{These authors contributed equally to this work.}
\abstract{
Existing multi-agent deep reinforcement learning (MADRL) methods for multi-UAV navigation face challenges in generalization, particularly when applied to unseen complex environments. 
To address these limitations, we propose a Dual-Transformer Encoder-based Proximal Policy Optimization (\textit{DTPPO}) method. 
DTPPO enhances multi-UAV collaboration through a Spatial Transformer, which models inter-agent dynamics, and a Temporal Transformer, which captures temporal dependencies to improve generalization across diverse environments. This architecture allows UAVs to navigate new, unseen environments without retraining. Extensive simulations demonstrate that DTPPO outperforms current MADRL methods in terms of transferability, obstacle avoidance, and navigation efficiency across environments with varying obstacle densities. 
The results confirm DTPPO's effectiveness as a robust solution for multi-UAV navigation in both known and unseen scenarios.
}

\keyword{Multi-UAV navigation; partially observable
Markov decision process; multi-agent deep reinforcement learning; cross-scenario transferability} 

\begin{document}

%%%%%%%%%%%%%%%%%%%%%%%%%%%%%%%%%%%%%%%%%%

% The order of the section titles is different for some journals. Please refer to the "Instructions for Authors” on the journal homepage.

\section{Introduction}
In recent years, unmanned aerial vehicles (UAVs) (also known as \textit{drones}) have rapidly emerged as vital tools in numerous applications, ranging from search and rescue missions to infrastructure monitoring and delivery services \cite{shakhatreh2019unmanned,mohsan2022towards}. 
However, the challenge of ensuring safe and efficient navigation in complex and dynamic environments, particularly when multiple UAVs are involved, remains an open problem. 
In multi-UAV scenarios, UAVs must coordinate their actions to avoid obstacles \cite{huang2019collision}, maintain efficient paths \cite{bellingham2002cooperative}, and successfully complete their missions in environments with limited or partially observable information.
Various centralized-based multi-UAV navigation systems have been developed to address these challenges \cite{lewis2014cooperative, liu2020prediction, liu2023graph}.
A central server manages all UAVs' actions by leveraging global information about their states and observations. This global control can guarantee safety and near-optimal path planning under ideal conditions, as it allows for complete knowledge of the environment and inter-drone interactions.
However, centralized systems face significant limitations, such as the high reliance on stable communication with a central server and the escalating computational burden as the number of UAVs increases, making them less scalable and vulnerable to failures if the server is compromised.

Compared to the centralized methods, some traditional decentralized multi-UAV navigation systems \cite{van2010optimal,van2011reciprocal}, such as those based on the velocity obstacle framework, allow agents to make independent decisions while avoiding collisions \cite{snape2011hybrid,douthwaite2019velocity}. 
However, these methods often require extensive communication between agents and are highly sensitive to environmental interference, making them difficult to implement in real-world scenarios. 
Moreover, such approaches rely on complex parameter tuning, limiting their generalization. 
To overcome these limitations, our work focuses on distributed control using Multi-Agent Deep Reinforcement Learning (MADRL) algorithms \cite{gronauer2022multi}, which allows UAVs to learn cooperative strategies in dynamic, uncertain environments without the need for constant communication or predefined rules.

Existing MADRL-based methods have shown promise in addressing the challenges of multi-UAV navigation \cite{bouhamed2020autonomous, qie2019joint, rybchak2024comparative, yu2022surprising}. 
These approaches model the problem as decentralized partially observable Markov decision processes (Dec-POMDPs) and apply deep reinforcement learning to train agents to make decisions based on their limited perception. 
Typical methods like multi-agent deep deterministic policy gradient (MADDPG) \cite{lowe2017multi} have been successfully applied to tasks such as formation control and obstacle avoidance, but they struggle with issues such as non-stability during training and limited generalization to more complex environments. 
Recent approaches based on recurrent deterministic policy gradient (RDPG) \cite{xue2023multi} and proximal policy optimization (PPO) \cite{hodge2021deep} applied for multi-UAV navigation tasks have demonstrated advantages in handling partial observation and improving training stability, respectively.
Despite these advancements, the trained models often face significant limitations when applied to new, unseen environments. Current methods typically require retraining in each new scenario, leading to substantial computational costs and rendering them impractical for real-time applications.

\begin{figure}
\centering
\includegraphics[width=1.0\linewidth]{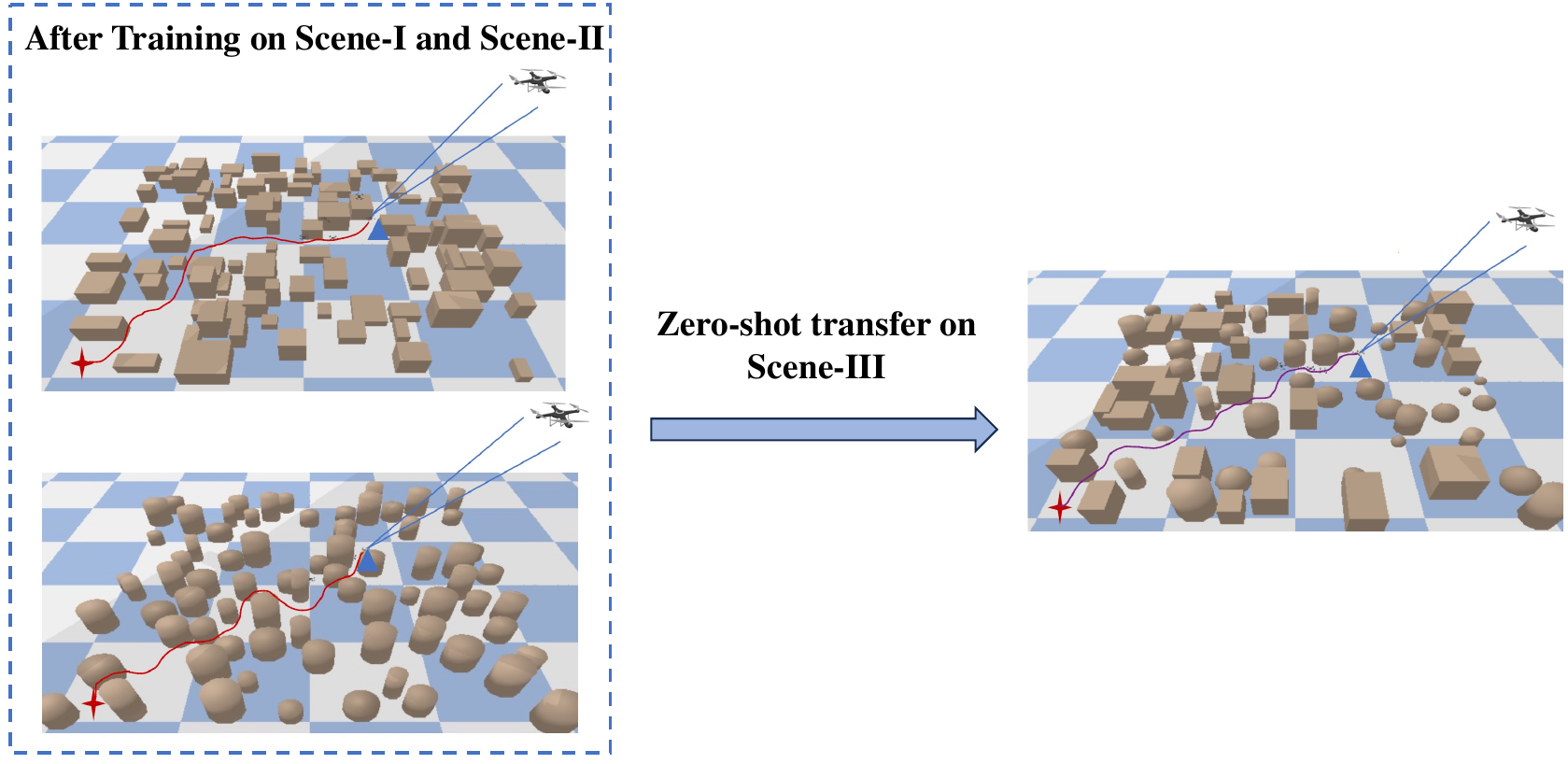}
\caption{A schematic illustration of zero-shot transfer to a previously unseen environment (Scene-III) after training on known environments (Scene-I and Scene-II). \label{cross_sce}}
\end{figure}

To address this issue, we propose a Dual-Transformer Encoder based Proximal Policy Optimization (\textit{DTPPO}) method, which enables multi-UAV systems to transfer learned knowledge from known scenarios to new, unseen environments without the need for extensive retraining (as shown in Figure~\ref{cross_sce}). 
Our approach incorporates two key components: (1) a Spatial Transformer, which enhances collaboration between neighboring UAVs by modeling the inter-agent dynamics, and (2) a Temporal Transformer, which captures the temporal evolution of multi-UAV trajectories across various environments. This Dual-Transformer (Dual-T) architecture is explicitly designed to improve transferability across diverse environments with different obstacle densities and configurations.
Through co-training across multiple scenarios, DTPPO ensures that the learned policies generalize well beyond the training environments, enabling UAVs to adapt quickly to new environments without retraining. Furthermore, by leveraging the powerful PPO algorithm, DTPPO balances exploration and exploitation, allowing for robust policy optimization in challenging navigation tasks.

In summary, the main contributions of this paper are as follows:
\begin{itemize}
    \item We introduce a novel Dual-Transformer architecture for multi-UAV navigation that enhances inter-agent coordination through spatial and temporal modeling.
    \item We develop a co-training framework that allows UAVs to learn generalized navigation strategies across diverse environments with varying obstacle densities.
    \item We validate the effectiveness of DTPPO through extensive simulations, demonstrating superior performance and transferability compared to state-of-the-art MADRL-based methods.
\end{itemize}

The remainder of this paper is organized as follows: Section \ref{related works} reviews related work on multi-UAV navigation and deep reinforcement learning. 
Section \ref{Preliminary} provides the necessary background and prior knowledge related to our problem setup. Section \ref{Methodology} outlines the proposed methodology, including the Dual-Transformer Encoder and PPO-based multi-scenario co-training. 
Section \ref{Experiment} details the experimental setup and results, and Section \ref{Conclusions} concludes the paper with insights and future directions.

%%%%%%%%%%%%%%%%%%%%%%%%%%%%%%%%%%%%%%%%%%
\section{Related Works}
\label{related works}
% \subsection{Deep Reinforcement Learning Based UAV Navigation}
In this section, we review the existing works on multi-UAV Navigation with regards to deep reinforcement learning algorithms.
In recent years, as Deep Reinforcement Learning (DRL) has achieved great success in many control tasks, such as traffic control \cite{lu2024dualight,mao2024pdit,mao2022transformer,du2024felight,jiang2024gesa,jiang2024x,ruan2024coslight,jiang2024guidelight}. In the past two years, Large Language Model (LLM)-based agents \cite{ruan2023tptu,kong2024tptu,zhang2024controlling} have also emerged. In the application of UAV, DRL is integrated to achieve UAV autonomous navigation and enhance real-time decision-making capabilities.
Wang et al. \cite{wang2019autonomous} formulated the navigation problem as a partially observable Markov decision process (POMDP), and employed an online DRL method to solve it.
In work \cite{pham2018reinforcement}, a function
approximation based RL algorithm was presented to deal with a large number of state representations and to obtain faster convergence.
Li et al. \cite{li2022efficiency} designed a DRL-based UAV navigation framework, which considers temporal abstractions and chooses the frequency of action decisions dynamically with efficiency regularization.
To assist multiple UAVs in reaching their goal points without obstacle collision in unknown complex environments, many multi-agent DRL (MADRL) algorithms can be utilized to learn the optimal trajectory for each drone.
In multi-UAV navigation, multi-agent Deep Deterministic Policy Gradient (MADDPG) \cite{lowe2017multi} methods have been applied extensively to address complex tasks such as formation control, collaborative target tracking, and obstacle avoidance in dynamic environments \cite{bouhamed2020autonomous,he2020deep,qie2019joint}.
The work \cite{qie2019joint} leveraged MADDPG to solve target assignment and path planning simultaneously.
To boost learning effects in unstable 3D environments, Xue et al. \cite{xue2023multi} proposed a multi-agent Recurrent Deterministic Policy Gradient (MARDPG) algorithm for developing navigation policy for multi-UAV.
While these DPG-based methods excel in handling continuous action spaces and multi-UAV coordination, Proximal Policy Optimization (PPO) based methods have also gained significant attention in UAV navigation due to the robustness and ability to balance exploration and exploitation during policy optimization \cite{rybchak2024comparative}.
Multi-agent PPO (MAPPO) \cite{yu2022surprising} can be applied in multi-UAV systems, enabling each UAV to learn its own policy while still benefiting from centralized training.
Hodge et al. \cite{hodge2021deep} developed an adaptive navigation framework using MAPPO combined with incremental course learning, allowing UAVs to efficiently track targets using real-time sensor data.
To tackle the challenge of exploring unknown complex environments, Moltajaei et al. \cite{moltajaei2024policy} employed on-policy RL with MAPPO to guide multiple UAVs in exploring areas of interest. 
Additionally, Chikhaoui et al. \cite{chikhaoui2022ppo} integrated energy constraints into a MAPPO-based DRL framework, enhancing UAV efficiency and extending operational duration.

Although the aforementioned MADRL methods enable UAVs to learn efficient navigation strategies in complex and dynamic environments, they are environment-specific (in other words, training and testing must be conducted in the same environment). 
Even if UAVs are trained using MADRL algorithms across multiple different maps or environments to learn a general navigation strategy, their performance remains limited in unseen environments. 
Therefore, this study aims to achieve strong generalization performance by coordinating multiple UAVs across various environments. From a broader perspective, various techniques can potentially improve a model's generalizability and transfer to new unseen data or tasks, such as multi-task learning \cite{zhang2021survey,lan2023mm}, transfer learning \cite{pan2009survey,li2022profile}, meta-learning \cite{jiang2024x}, domain adaptation \cite{farahani2021brief,guo2024online}, contrastive learning \cite{chen2020simple,mao2022jointly,wang2023correlated,li2024non,li2024enabling}, and so on. In this work, we propose a dual-transformer-based meta-reinforcement learner.

% trained a quad-rotor to learn to navigate to the target point using a PID assisted Q-learning algorithm in an unknown environment.
% considerable progress has been made in the fields of UAV autonomous navigation tasks, driven by advances in Deep Reinforcement Learning (RL) methods. 

% To minimize the dependence of obstacle avoidance algorithms on specific environment settings , Reinforcement Learning (RL) has been integrated to achieve UAV autonomous navigation.
% Pham et al [7] trained a quad-rotor to learn to navigate to the target point using a PID assisted Q-learning algorithm in an unknown environment.
% and enhance the applicability

%%%%%%%%%%%%%%%%%%%%%%%%%%%%%%%%%%%%%%%%%%
\section{Preliminary}
\label{Preliminary}
In this work, we study the multi-UAV navigation task across various complex and dynamic environments. We introduce the UAV system model and problem statement as follows.

\subsection{UAV System Model}
Referring to prior works \cite{xue2023multi,panerati2021learning}, we model the UAV as a quadrotor with a 12-dimensional state, which includes the absolute position $[x, y, z]$ of the UAV in the world coordinate frame, the Euler angles $[\phi, \theta, \psi]$ representing the UAV’s rotation state, the velocity $[v_x, v_y, v_z]$ along the three axes of the coordinate frame, and the angular velocity $[\omega_x, \omega_y, \omega_z]$.
Thus, the complete state vector $\boldsymbol{s}$ can be expressed as $s = [x, y, z, \phi, \theta, \psi,v_x, v_y, v_z,\omega_x, \omega_y, \omega_z]$.
% \begin{equation}
% \end{equation}
The state $s$ of a UAV captures both its position and orientation in the 3D space.
To control the UAV, we utilize a 4-dimensional velocity vector as the control action $a = [v_x, v_y, v_z, v_M]$, where $v_x$, $v_y$, and $v_z$ are the components of a unit vector representing the direction of motion in the 3D space, and $v_M$ denotes the magnitude of the desired velocity.
Thus, the control action $a$ can specify the direction and speed at which the UAV should move. 

To successfully reach the designated target point without colliding with obstacles in the environment, MADRL will be applied to control multi-UAV navigation in complex environments. 
During navigation, environmental information is collected in real-time by the UAV's sensors, and corresponding action controls are made. After executing the actions, the UAV transitions to a new state and receives feedback from the environment. 
Using this feedback, the UAV can update its action selection strategy, enabling it to reach the target more efficiently while avoiding obstacles in the environment.
In this paper, we aim to design a MADRL algorithm that enables multiple UAVs to learn general and effective action strategies for navigation tasks, even in different complex environments, such as those with varying terrains or obstacle densities.

\subsection{Problem Statement}
The problem of multi-agent UAV action control in various scenarios can be formulated as the Decentralized Partially Observed Markov Decision Processes (Dec-POMDPs) \cite{bernstein2002complexity}.
The goal for multiple UAVs is to cooperate and navigate safely through each scenario while avoiding obstacles and efficiently reaching their target destinations.
Given a set of environments $E$ with different types of obstacles and obstacle densities, each agent $i$ controls a drone $D_i$ in an environment $e \in E$. 
We consider the top $n$ nearest neighboring drones $\mathbf{D}_{\mathcal{N}_{i}}$ of drone $D_i$ within its sensing range, where $\mathcal{N}_{i} = \{\mathcal{N}_{1}, ..., \mathcal{N}_{n}\}$.

Then, we represent this POMDP using the tuple $<\mathcal{S}, \mathcal{O}, \mathcal{A}, \mathcal{P}, r, \gamma>$, where $\mathcal{S}$ is the state space and $s_t \in \mathcal{S}$ denotes the state of all drones at time step $t$. 
The local observation can be obtained through an observation function $D(s): \mathcal{S} \rightarrow \mathcal{O}$.
$\mathcal{A}$ denotes the action space for each agent.
When $m$ agents take a joint control actions $\mathbf{a}_t = \{a_t^{1}, ..., a_t^{m}\}$ in the environment $e$, the state transition $\mathcal{P}(s_{t+1}|s_{t},\mathbf{a}_t) = \mathcal{S} \times \mathcal{A} \rightarrow \mathcal{S}$ occurs and each agent $i$ obtained a reward $r_t^{i}$.
Due to the limited sensing range of the drone, the environment is partially observed, and each agent $i$ can only have access to the joint actions $a_{t}^{i, \mathcal{N}_{i}}$ and the local observation $o_{t+1}^{i, \mathcal{N}_{i}}$, which 
respectively include the local control actions and state transitions of the target drone $D_i$ and its top $n$ nearest neighboring drones $\mathbf{D}_{\mathcal{N}_{i}}$.
Therefore, each agent gets $(o_{t+1}^{i, \mathcal{N}_{i}}, a_{t}^{i, \mathcal{N}_{i}}, r_t^{i})$ at the next time step $t+1$.
When updating the action policy, the cumulative reward for all agents in each scenario $\sum_t \sum_{m} \gamma_t r_t^{i}$ is expected to be maximized, where $\gamma$ denotes the discounted factor.

In this paper, we aim to develop a generalized multi-UAV navigation policy capable of performing well across various scenarios, even though these scenarios have not been encountered during training.
As shown in Figure \ref{map}, maps with different obstacle types and varying obstacle densities represent distinct environments. 
Our objective is to learn an action control policy parameterized by $\theta$, which can distinguish between tasks (i.e., learning on different environments) in the embedding space, and minimize the loss across these diverse tasks:
\begin{equation}
    \theta = \arg \min_{\theta} \frac{1}{m |E|} \sum_{e \in E} \sum_{i=1}^{m} \mathcal{L}(f_{\theta}(D_i), D_i),
\end{equation}
where $\theta$ represents the policy parameter, $f_{\theta}(D_i)$ is the control action output for UAV $D_i$, which denotes the policy to solve navigation task in environment $e$.

%%%%%%%%%%%%%%%%%%%%%%%%%%%%%%%%%%%%%%%%%%

\section{Methodology}
\label{Methodology}
In this section, we present a general MADRL method for cross-scenario multi-UAV navigation task, referred to as \textit{DTPPO}.
We first provide an overview of our method, followed by the introduction of the Dual-Transformer (Dual-T) Encoder module, which is composed of the Spatial Transformer and the Temporal Transformer.
Additionally, we illustrate the details of the co-training process across diverse scenarios using the PPO algorithm.

\begin{figure}
\centering
\includegraphics[width=1.0\linewidth]{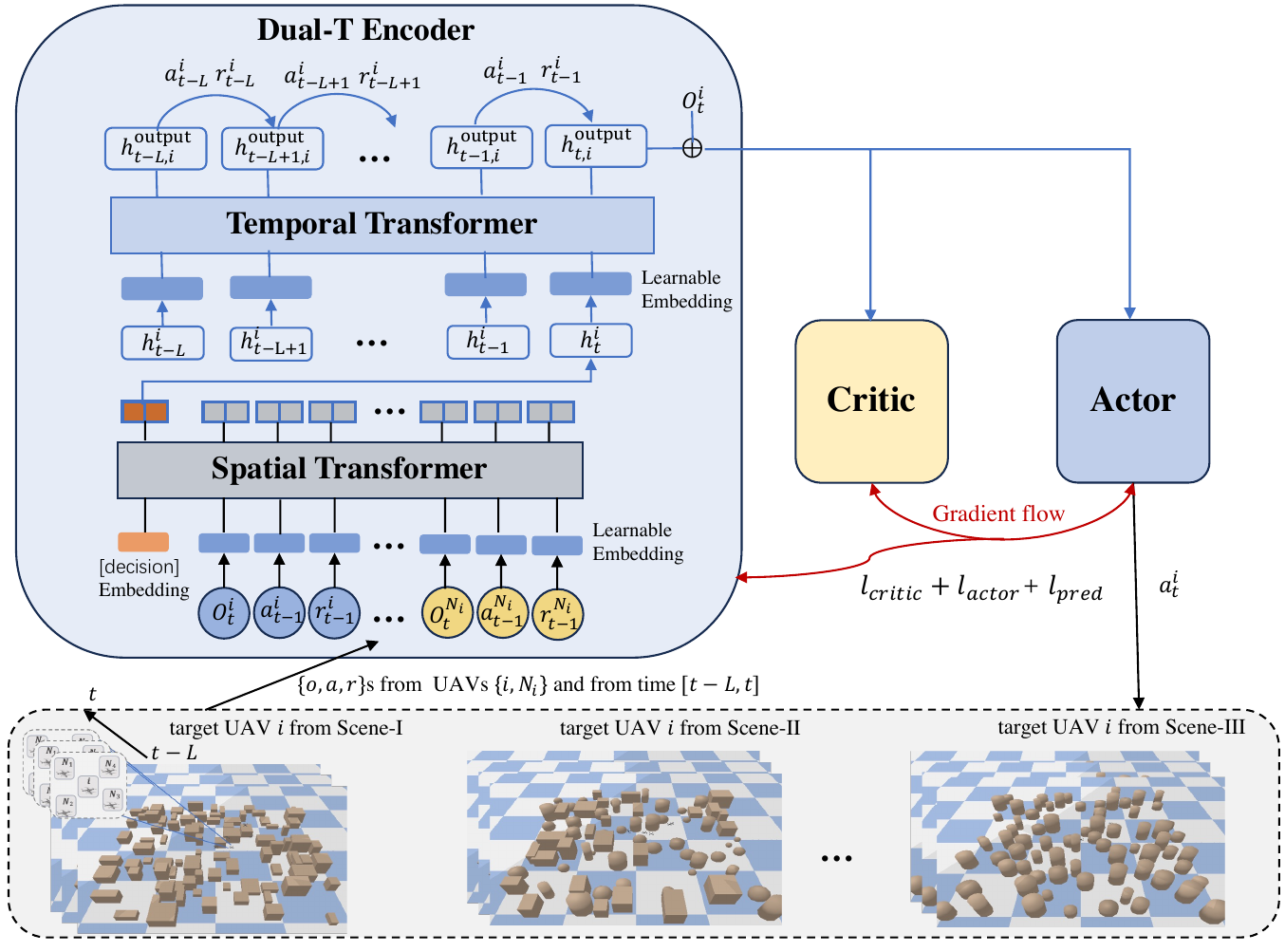}
\caption{Overview of DTPPO. \label{overview:dtppo}}
\end{figure}  

\subsection{Overview of DTPPO}
The overall training process of our method is shown in Figure \ref{overview:dtppo}. 
DTPPO is trained using UAVs' MDP trajectories across multiple environments within a batch. 
Specifically, for a target agent $i$ and its neighboring agents $\mathcal{N}_{i}$, their MDP trajectories $(o, a, r)$ in a certain range of time steps $[t-L, t]$ are sampled and fed into the Dual-T Encoder module, where $L$ denotes the length of the time frame. 
The Dual-T Encoder is composed of two transformers: the Spatial Transformer and the Temporal Transformer. At time step $t$, the Spatial Transformer takes the MDP information of each UAV and its neighboring UAVs as input, enhancing the collaboration between agents within the UAV's sensing range. 
The Temporal Transformer utilizes historical MDP trajectories as context to infer the current task, thereby improving transferability.

Referring to previous work \cite{panerati2021learning,xue2023multi}, four types of kinematic information are selected from the observations as states: absolute position, Euler angles, velocity, and angular velocity.
Each UAV utilizes a 4-dimensional velocity vector as its control action to execute the next movement.
The full observation for each agent $o^{i} = o^{i, \mathcal{N}_{i}}$ contains the local observations from the target agent $i$ and its neighbors. The local observation consists of the current state information concatenated by historical actions during the last $\Delta t$ time steps, where we set $\Delta t = 15$. 
The reward $r$ can be defined as the weighted sum of three components: transfer reward, collision penalty, and free space reward. 
The transfer reward is denoted as follows.
\begin{equation}
        r_{trans} =  \left[ \left( \|\mathbf{x}_{\text{target}} - \mathbf{x}_{t}\|_2 - \|\mathbf{x}_{\text{target}} - \mathbf{x}_{t-1}\|_2 \right) + \max\left( 0, \left( 2 - \|\mathbf{x}_{\text{target}} - \mathbf{x}_{t}\|_2^2 \right) \right) \right]
\end{equation}
The first term in the function $r_{trans}$ measures the change in distance to the target between consecutive time steps, and the second term ensures that if the UAV gets very close to the target (i.e., within a distance of 2 units), it receives an additional positive reward. 
Thus, the transfer reward encourages the UAV to approach its target efficiently while avoiding unnecessary detours. 
Combined the collision penalty $r_{col}$ (we set to $-1.0$) and free space reward $r_{free}$ (we set to $0.04$), which encourage UAV to explore toward a safe space, we define the total reward function as:
\begin{equation}
    r_{total} = \lambda_1 r_{trans} + \lambda_2 r_{col} + \lambda_3 r_{free}
\end{equation}
where $\lambda_1$, $\lambda_2$, $\lambda_3$ are scale factors.

\subsection{Dual-Transformer Encoder Module}
The Dual-Transformer Encoder (Dual-T Encoder) module is the core of the DTPPO algorithm, designed to handle both the spatial collaboration between UAVs and the temporal dynamics of their MDP trajectories across various environments. 
It includes the Spatial Transformer and the Temporal Transformer, working together to process the transition for each agent.

\subsubsection{Spatial Transformer}
The Spatial Transformer is designed for enhancing collaboration between the target UAV $D_i$ and its top $n$ nearest neighbors $\mathbf{D}_{\mathcal{N}^i}$ within the sensing range. 
Here, we set $n=4$.
At each time step $t$, the Spatial Transformer has access to the MDP features, including the current observations $o_t$, the previous actions $a_{t-1}$, and the rewards $r_{t-1}$ from both the target UAV and its neighboring UAVs.
Unlike previous MADRL-based navigation methods \cite{wei2024uav,wu2022improved}, which only consider the states of neighboring drones for cooperation, Spatial Transformer considers the complex interrelations among neighboring drones' observations, actions, and rewards.
Regardless of the type of map, UAVs share a common characteristic: within the group of $n+1$ closely located UAVs, one's action will affect another one's navigation route decisions. 
Therefore, in the policy learning process, considering only the states is inadequate for capturing the mutual influence between neighboring UAVs, which can further exacerbate instability during the co-training process across various scenarios \cite{jiang2024x,zang2020metalight}.

In the Spatial Transformer, for each UAV, we leverage the full MDP features $\mathbf{m}_t^i$ from the target drone $i$ and its neighbors to boost up the coordination during navigation. 
As DTPPO is an online RL algorithm, only the current observation $o_t$, the previous action, and reward $a_{t-1}$, $r_{t-1}$ can be acquired.
The concatenated MDP features of agent $i$ can be expressed as $\mathbf{m}_t^i = [\mathbf{o}^{i}_{t}, \mathbf{a}^{i}_{t-1}, \mathbf{r}^{i}_{t-1}]$.
As $\mathbf{o}^{i}_{t}$, $\mathbf{a}^{i}_{t-1}$, $\mathbf{r}^{i}_{t-1}$ have different dimensions, they can be passed through three different learnable linear projections $\mathbf{W} = [\mathbf{W}_o$, $\mathbf{W}_a$, $\mathbf{W}_r]$, allowing them to be transformed into a common $d$-dimensional latent space:
\begin{equation}
    \mathbf{m}_t^i \mathbf{W} = \left[\mathbf{o}^{i}_{t} \mathbf{W}_o, \mathbf{a}^{i}_{t-1} \mathbf{W}_a,\mathbf{r}^{i}_{t-1}\mathbf{W}_r\right] \in \mathbb{R}^{3 \times d}.
\end{equation}

By concatenating neighboring agents, the full MDP transition of agent $i$ at time step $t$ can be defined as:
\begin{equation}
    \mathbf{M}_t^i = \left[\mathbf{m}_t^i \mathbf{W}; \mathbf{m}_t^{\mathcal{N}_{1}} \mathbf{W}; ...; \mathbf{m}_t^{\mathcal{N}_{n}} \mathbf{W} \right] \in \mathbb{R}^{3 (n+1) \times d}
\end{equation}
The resulting embedding $\mathbf{M}_t^i $ encapsulates the spatial relationships and cooperation among UAVs, which is essential for effective multi-UAV collaboration, especially in densely populated or obstacle-rich environments.
Notely, when the number of neighbors is fewer than $n$, we apply zero-padding and include a binary indicator embedding to $\mathbf{o}_t$ and $\mathbf{a}_{t-1}$, to indicate whether the neighboring drone exists.

Referring to the works \cite{dosovitskiy2020image, jiang2024x}, we prepend a learnable $\texttt{[decision]}$ token $\mathbf{q}_{\text{decision}}$, so that the state at the output of the Spatial Transformer can be served as the drone's representation $\mathbf{d}_t$.
Moreover, standard positional embedding $\mathbf{E}_{pos}^{\text{S}} \in \mathbb{R}^{3 (n+1) \times d}$ is added to each input token to retain positional information \cite{vaswani2017attention}, and the input to the Spatial Transformer at time step $t$ is given by:
\begin{equation}
    \mathbf{z}_{t,i}^{\text{S}} = \left[\mathbf{q}_{\text{decision}}; \mathbf{m}_t^i \mathbf{W}; \mathbf{m}_t^{\mathcal{N}_{1}} \mathbf{W}; ...; \mathbf{m}_t^{\mathcal{N}_{n}} \mathbf{W} \right] + \mathbf{E}_{pos}^{\text{S}}.
\end{equation}
Then we feed $\mathbf{z}^{\text{S}}_{t, i}$ to the Spatial Transformer with multi-head self-attention layers, and obtain a drone's embedding $\mathbf{h}_t^i = \textbf{SpatialTransformer}(\mathbf{z}_{t, i}^{\text{S}})$.

\subsubsection{Temporal Transformer}
The Temporal Transformer plays a crucial role in ensuring that the model generalizes well to unseen environments by capturing long-term temporal dependencies.
It processes the sequence of embeddings $\mathbf{h}_{[t-L:t]}^i$ generated by the Spatial Transformer over the last $L$ time steps, utilizing multi-head self-attention to extract temporal relationships. 
Thus, DTPPO is a context-based MADRL method.

At each time step $t$, the Temporal Transformer takes as input the spatial embeddings $\mathbf{h}_{[t-L:t]}^i$ for agent $i$ obtained from the Spatial Transformer, which is first projected to a lower-dimensional space using a trainable projection matrix $W'$. 
These projections encode the relevant spatial and temporal features, enabling the Temporal Transformer to capture the task-related dynamics over time steps.
Similarly, we add the positional embedding $\mathbf{E}^{\text{T}}_{pos} \in \mathbb{R}^{L \times d'}$ (where $d'$ denotes the lower dimensionality) to retain the sequential order of the input. 
The input to the Temporal Transformer for the time window $[t-L, t]$ is:
\begin{equation}
    \mathbf{z}_{[t-L:t],i}^{\text{T}} = \left[\mathbf{h}_{t-L}^i \mathbf{W'}; \mathbf{h}_{t-L+1}^{i} \mathbf{W'}; ...; \mathbf{h}_t^{i} \mathbf{W'} \right] + \mathbf{E}_{pos}^{\text{T}}.
\end{equation}

Then, the Temporal Transformer operates over the input within dimensions $\mathbb{R}^{L \times L}$ using the attention mechanism, which consists of six multi-head self-attention layers.
Thus it can capture the evolving environmental dynamics related to UAVs by leveraging historical data (i.e., MDP trajectories), and extract meaningful patterns for the UAV's next control actionsover time.
The output of  the Temporal Transformer can be defined as:
\begin{equation}
    \mathbf{h}_{[t-L:t],i}^{\text{output}} = \textbf{TemporalTransformer}(\mathbf{z}_{[t-L:t],i}^{\text{T}}).
\end{equation}

To further enhance the UAV's understanding of environmental dynamics, we introduce a dynamic predictor between the output of the Temporal Transformer at each time step.
This dynamic predictor performs autoregressive prediction, which encourages the Temporal Transformer to model the cross-scenario dynamics effectively.
Specifically, the predictor works by taking the output at the previous time step $\mathbf{h}^{\text{output}}_{t-1,i}$ and concatenating it with the joint actions $\mathbf{a}_{t-1}^{i, \mathcal{N}_i}$ and rewards $\mathbf{r}_{t-1}^{i, \mathcal{N}i}$ from the target UAV and its neighbors. The goal is to predict the next temporal embedding $\mathbf{\hat{h}}^{\text{output}}_{t,i}$ using a single-layer MLP:
\begin{equation}
   \mathbf{\hat{h}}^{\text{output}}_{t,i}  = \mathbf{MLP} \left(\left[\mathbf{h}^{\text{output}}_{t-1,i}, \mathbf{a}_{t-1}^{i, \mathcal{N}_i},  \mathbf{r}_{t-1}^{i, \mathcal{N}i}  \right]  \right)
\end{equation}
The training objective of the dynamic predictor is to minimize the prediction loss $l_{pred} = \mathbf{MSE}(\mathbf{\hat{h}}^{\text{output}}_{t,i}, \mathbf{h}^{\text{output}}_{t-1,i})$, defined as the mean squared error (MSE) between the predicted embedding $\mathbf{\hat{h}}^{\text{output}}_{t,i}$ and the actual output embedding $\mathbf{h}^{\text{output}}_{t-1,i}$ of the Temporal Transformer.

\subsection{PPO-based Co-Training on Various Scenarios}
To learn the decision policy, the output of the Dual-T Encoder is used as input to the Actor-Critic framework in the PPO algorithm \cite{schulman2017proximal}.
Specifically, both the Actor and Critic networks are implemented as two-layer MLPs, where the Actor generates the control actions for the UAV, and the Critic evaluates the state value to guide the learning process.
For the policy $\pi$, the actor-network takes $\mathbf{h}^{\text{output}}$ as input and makes the control action $\mathbf{a}_t^i$ for the target drone $i$.
In addition, we implement a \textit{residual link} to prevent over-abstraction of the agent's embedding via Dual-T Encoder. 
The residual connection adds direct self-observation $\mathbf{o}_t^i$ to the $\mathbf{h}_{t,i}^{\text{output}}$, ensuring that the actor has both a high-level abstract representation of the current environment and enough up-to-date observation information from the target drone $i$.
The actor network then outputs the action $\mathbf{a}_t^i$ using the policy $\pi$ as follows:
\begin{equation}
\label{policy}
    \mathbf{a}_t^i \sim \pi(\cdot \mid \mathbf{h}_{t,i}^{\text{output}} + \mathbf{o}_t^i)
\end{equation}
In Eq.~\ref{policy}, $\mathbf{h}_{t,i}^{\text{output}}$ represents the high-level feature embedding generated by the Dual-T Encoder. 
It provides a comprehensive context for decision-making within the dynamic and complex environment, also enhancing generalization across diverse scenarios.
Conversely, $\mathbf{o}t^i$ represents the self-observation of the target UAV, focusing on its current state.
This is critical for making precise, real-time adjustments in response to sudden environmental changes.
Thus, combined with prediction loss $l_{pred}$, the overall optimization objective function can be written as:
\begin{equation}
\label{loss}
    l_{DTPPO} = \delta_1 l_{actor} + \delta_2 l_{critic} + \delta_3 l_{pred}
\end{equation}
where $\delta_1$, $\delta_2$, $\delta_3$ denote hyperparameters. The Actor loss $l_{actor}$ and Critic loss $l_{critic}$ can be referred to as the original PPO method \cite{schulman2017proximal}.
Finally, we can employ co-training across multiple scenarios to increase training data diversity for better model generality. The UAVs will be stochastically chosen from various scenarios within each training batch.
In these scenarios, there are obstacles and structures of various shapes or obstacle densities, which correspond to navigation tasks following different task distributions.
This setup encourages the agent to learn more generalized knowledge while enabling a stable learning process.
The training process of DTPPO can be summarized in Algorithm~\ref{al:dtppo}.

\begin{algorithm}[ht]
    \caption{Training process of DTPPO}
    \label{al:dtppo}
    \textbf{Input:}  A set of target UAVs $\mathcal{D}$ from various scenarios $\mathcal{S}$, training episodes $E$, the number of neighbors $n$, the input length $L$ for the Temporal Transformer, the PPO epochs $Epoch$. \\ 
    \textbf{Initialize:} MDP buffer $\mathcal{D}$, policy parameters $\theta$.
    \begin{algorithmic}[1]
        \FOR{episode = 1 to $E$}
            \STATE Initialize buffer $\mathcal{D} \leftarrow \emptyset$ 
            \FOR{each scenario $s \in \mathcal{S}$  in parallel} 
                \STATE Use the top $n$ nearest neighbors $\mathcal{N}_i$ for each UAV $i$
                \FOR{each time step $t$}
                    \STATE Retrieve the last $L$ transitions $\{ \mathbf{m}^i_{t-l} \}^{L}_{l=0}$ for each UAV and add to buffer $\mathcal{D}$
                    \STATE Make action $\mathbf{a}_t^i$ using policy $\pi^\theta$ according to Eq. \ref{policy}, and take joint action $\{ \mathbf{a}_t^1, ..., \mathbf{a}^n \}$
                    \STATE Observe the next state $\mathbf{o}_{t+1}^i$ and current reward $\mathbf{r}_t^i$
                \ENDFOR
            \ENDFOR
            \FOR{e = 1 to $Epoch$}
                \STATE Sample mini-batch data from buffer $\mathcal{D}$
                \STATE Calculate dynamic predictions $\{ \mathbf{\hat{h}}^{\text{output}}_{t-l,i} \}^{L-1}_{l=0}$.
                \STATE Compute the total loss $\mathcal{L}$ using Eq. \ref{loss} and update policy parameters $\theta$
            \ENDFOR
        \ENDFOR
        \RETURN Optimized policy $\pi^\theta$
    \end{algorithmic}
\end{algorithm}

%%%%%%%%%%%%%%%%%%%%%%%%%%%%%%%%%%%%%%%%%%
\section{Experiment}
\label{Experiment}
\subsection{Experiment and Parameter Setting}
\begin{figure}[H]
% \begin{adjustwidth}{-\extralength}{0cm}
\centering
\includegraphics[width=1.0\linewidth]{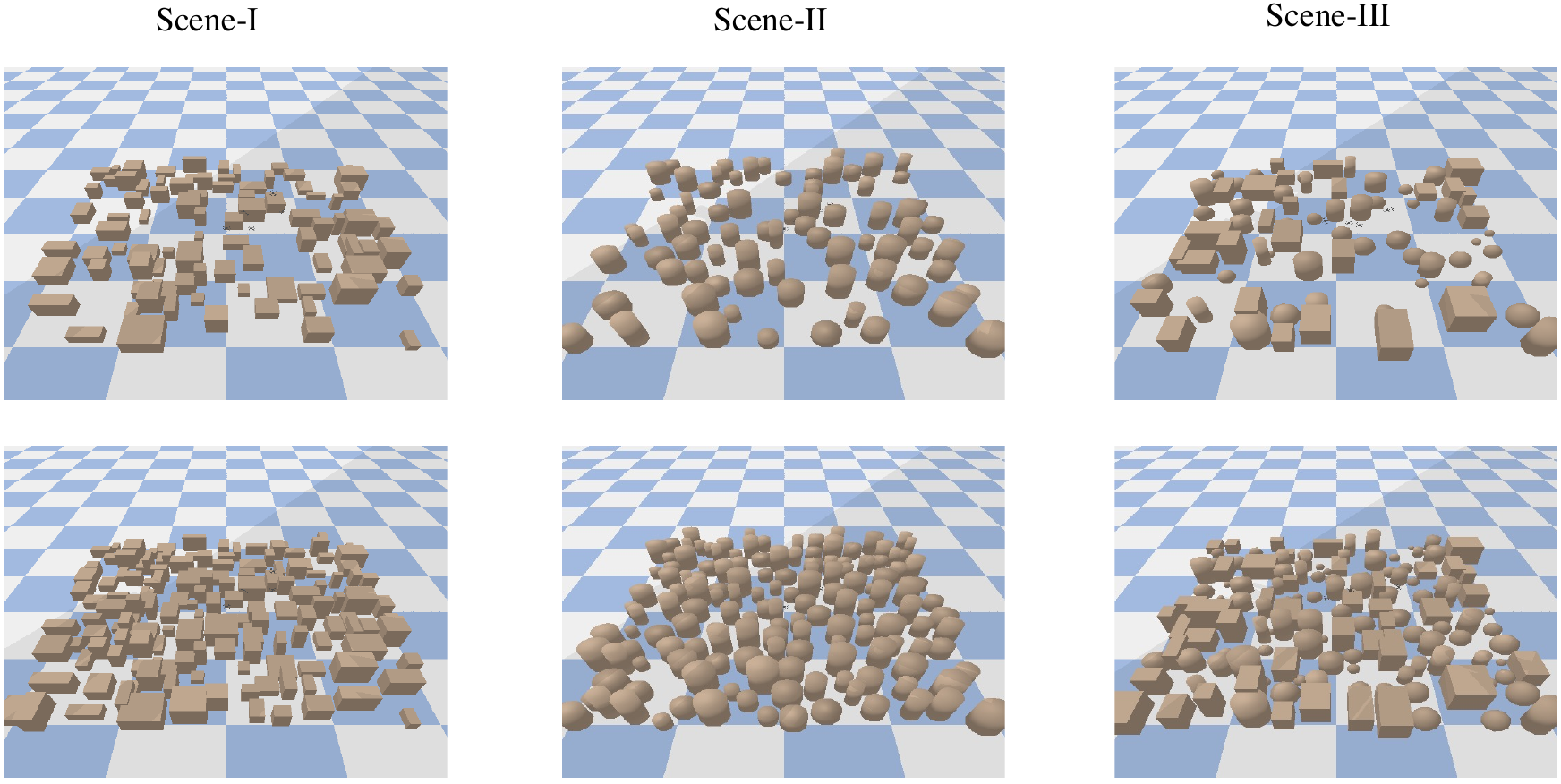}
% \end{adjustwidth}
\caption{The Navigation algorithm will be tested in the three types of environments: a square column obstacle, a cylindrical obstacle, and mixed obstacles. Different obstacle densities can be set for training. \label{map}}
\end{figure}

We utilize the simulated environment \textit{gym-pybullet-drones} \cite{panerati2021learning}, which supports the random generation of maps. The environment includes three types of obstacles: square pillars, cylinders, and mixed 3D obstacles. We refer to these environments as \textit{Scene-I}, \textit{Scene-II}, and \textit{Scene-III}, respectively. These settings are designed to replicate real-world obstacles, such as urban buildings and varying terrain features.
During training, the UAV agents navigate through these randomly generated environments. Obstacle density is defined as the percentage of space within the environment occupied by obstacles, with higher densities posing a greater challenge for UAV navigation. We use obstacle densities of $[10\%, 25\%, 50\%]$ for each type of map, resulting in a total of nine different maps for multi-scenario co-training. This setup encourages generalization across diverse obstacle distributions and task settings.
During evaluation, we use six generated unseen maps (as shown in Figure~\ref{map}) for testing our method.

The altitude of the UAVs is limited to the range [0.0 m, 30.0 m]. The control signal was normalized to the range [-1, 1] for stability. 
The reward function parameters are set as follows: transfer reward coefficient $\lambda_1 = 0.45$, collision penalty coefficient $\lambda_2 = 0.30$, and free space reward coefficient $\lambda_3 = 0.25$. 
The exploration reward is set to $r_{\text{free}} = 0.04$ and collision penalty is set to $r_{\text{col}} = -1.0$. 
When training our method, the hyperparameters used in the model are carefully tuned based on preliminary experiments to achieve optimal performance. The details of hyperparameters are listed in the Table~\ref{tb:detail}
All simulations are run on an Ubuntu 20.04 system with 32 GB RAM and a Tesla V100 GPU.
The UAVs are trained for a total of 1,000,000 episodes across multiple environments, which required approximately 38 hours to complete.
\begin{table}[H]
    \centering
    \caption{Implementation details of DTPPO. \label{tb:detail}}
    \begin{tabular}{|l|l|}
        \hline
        \textbf{Hyperparameters} & \textbf{Details} \\ \hline
        Learning rate & 5e-4 \\ \hline
        Actor loss coefficient $\delta_1$ & 1 \\ \hline
        Critic loss coefficient $\delta_2$ & 1 \\ \hline
        Dynamic predictor loss coefficient $\delta_3$ & 1e-2 \\ \hline
        Entropy coefficient & 1e-2 \\ \hline
        Discount factor $\gamma$ & 0.99 \\ \hline
        Clipping $\epsilon$ & 0.2 \\ \hline
        Number of Spatial transformer layers & 3 \\ \hline
        Number of Spatial transformer heads & 6 \\ \hline
        Number of Temporal transformer layers & 3 \\ \hline
        Number of Temporal transformer heads & 6 \\ \hline
        Spatial transformer embedding dimension  & 149 \\ \hline
        Temporal transformer embedding dimension  & 149 \\ \hline
        Temporal transformer horizon $L$ & 20 \\ \hline
        The number of neighbor drones $n$ & 4 \\ \hline
    \end{tabular}
\end{table}

\subsection{Baselines}
The proposed DTPPO will be compared to the following baseline methods to evaluate its effectiveness. 
The same states, actions, and rewards are applied in all baselines.
\begin{itemize}
    \item \textit{MADDPG} uses feedforward neural networks for learning. In MADDPG, the UAVs are trained in a centralized manner but execute their learned policies independently (decentralized execution). This method addresses the challenges of non-stationarity in multi-agent environments and reduces the variance in training across multiple UAVs.
    
    \item \textit{MARDPG} extends RDPG to the multi-agent deep reinforcement learning settings.  In MARDPG, each UAV perceives all other UAVs as part of the environment, without direct communication or cooperation between them. This can be referred to as Ind-MARDPG, where each UAV’s navigation policy is trained using a recurrent deterministic policy gradient. The UAVs in the environment adopt the same policy independently, without any exchange of information between agents.
    
    \item \textit{MAPPO} is an extension of the single-agent PPO algorithm to multi-agent systems. It combines centralized training with decentralized execution, where each UAV learns its own policy but benefits from joint learning with other agents. MAPPO offers more stable learning through the PPO clipping mechanism, which helps to avoid large policy updates. This makes MAPPO particularly suited for complex, dynamic environments where cooperation between agents is crucial.
\end{itemize}

\subsection{Evaluation Metrics}
To evaluate the performance of our proposed method, we utilize a set of quantitative metrics that capture the overall efficiency, safety, and robustness of the learned policies. The test metrics are presented as follows:

\begin{itemize}
    \item \textit{Average Transfer Reward}: This metric measures the average reward obtained by all UAVs during their navigation towards the target in different environments. 
    It reflects the efficiency of the learned navigation policies, with higher rewards indicating better performance in reaching the goal.

    \item \textit{Average Collision Penalty}: This metric records the average penalty incurred when any UAV collides with obstacles. It helps assess the safety of the navigation policies, with lower penalties indicating better obstacle avoidance and safer navigation.

    \item \textit{Average Free Space}: This metric evaluates how well the UAVs navigate through open, obstacle-free areas by averaging the rewards earned for doing so. 
    It indicates how effectively the UAVs avoid obstacles while maintaining efficient movement through less congested regions.
    
\end{itemize}

\begin{table}[!t] 
\caption{Test metrics on performing zero-shot transfer to various unseen scenes with different obstacle densities.\label{tab1}}
\begin{adjustwidth}{-\extralength}{0.5cm}
    \begin{tabularx}{\fulllength}{C|C|CCCCCC}
    \toprule
    \textbf{Metric}	 &\textbf{Method}	& Scene-I ($10\%$)	&Scene-I ($50\%$) & Scene-II ($10\%$)	& Scene-II ($50\%$) & Scene-III ($10\%$)	& Scene-III ($50\%$) \\
    \midrule
    \multirow{4}{*}{Avg. Transfer}  &MADDPG     &66.21  & 58.48   & 76.51  & 56.42   & 87.43	& 65.83  \\
    
   \multirow{4}{*}{Reward}  &MARDPG    &95.45  & 84.37   & 105.75  & 86.03   &92.32	&77.69 \\
    
   &MAPPO    & 168.39 & 151.58   & 196.85  & 148.57   & 166.43	& 134.90 \\
    
    &\textbf{DTPPO}	& \textbf{256.19} & \textbf{243.53}    & \textbf{239.26}  &\textbf{227.80}   & \textbf{231.26}  & \textbf{214.55} \\
    
    \midrule
    \multirow{4}{*}{Avg. Collision} 
    & MADDPG     &5.22  & 24.68  & 8.27  & 24.27   &13.66	& 33.25\\
    \multirow{4}{*}{Penalty} & MARDPG    &3.60  & 16.41  & 8.21  & 19.63  &10.25	& 28.26 \\
    & MAPPO     & 2.59 & 4.60  &3.24  &5.80   & 4.80 & 7.45 \\
    
    &\textbf{DTPPO}	& \textbf{1.20} & \textbf{1.61}  & \textbf{1.20}  & \textbf{2.56}   & \textbf{4.42} &  \textbf{5.58} \\
    \midrule
    
    \multirow{4}{*}{Avg. Free} 
    &MADDPG     &1.38  & 1.02  & 1.84  &0.46  &0.68	& 0.37\\
    
    \multirow{4}{*}{Space Reward} &MARDPG    &1.27  & 1.69 & 2.01  & 1.15  &1.28	& 0.68 \\
    
    &MAPPO     & 3.86 & 3.05  &3.02 &\textbf{4.80}   & 2.13 & 1.98 \\
    
    &\textbf{DTPPO}	& \textbf{4.65} & \textbf{3.97}  & \textbf{5.17}  & 4.56   & \textbf{3.41} &  \textbf{3.25} \\
    
    \bottomrule
    \end{tabularx}
\end{adjustwidth}
\end{table}

\subsection{Experimental Results}
In this section, we show the superior transferability and general great performance of DTPPO when performing navigation tasks on different unseen scenarios after training.

\subsubsection{Transferability on the Unseen Scenario}
We evaluate the transferability of DTPPO using a zero-shot setting, where the model is trained on several scenarios and then directly tested on unseen scenarios.
As shown in Table~\ref{tab1}, each column of results shows the performance of transferring to a new, unseen scenario after training on the preset nine scenarios.
The results clearly demonstrate that DTPPO achieves the best transfer performance in all tested scenarios compared to the other baseline methods.

\textbf{Cooperation is Key.} Our results highlight the importance of cooperation between UAVs for better transferability. DTPPO, by leveraging its Dual-Transformer architecture, enables efficient coordination among neighboring agents, which significantly improves navigation in unseen environments. This is particularly evident when compared to the baseline MADDPG, which does not model inter-agent collaboration to the same extent.\\
\textbf{Generalization to High-Density Obstacle Scenarios}. DTPPO excels in high-density obstacle scenarios, where the complexity of navigation increases substantially. For example, in Scene-III with 50\% obstacle density, DTPPO achieves a transfer reward of 214.55, far surpassing other methods like MAPPO (134.90) and MARDPG (77.69). This indicates that our model is able to generalize well even in challenging environments by learning more robust policies during training.\\
\textbf{Lower Collision Rates.} In addition to higher transfer rewards, DTPPO maintains lower collision penalties across all scenarios. In Scene-II with 50\% obstacle density, DTPPO achieves a collision penalty of only 2.56, which is significantly lower than MAPPO (5.80) and MADDPG (24.27). This demonstrates that DTPPO's learned policies are effective in avoiding obstacles while navigating through unseen environments.\\
\textbf{Efficient Use of Free Space.} DTPPO also makes better use of available free space in the environment, as evidenced by the higher Avg. Free Space Reward. In Scene-II with 10\% obstacle density, DTPPO achieves a reward of 5.17, outperforming all other baselines. This suggests that the model can efficiently navigate and utilize free areas, improving its overall navigation performance in novel environments.

Thus, DTPPO shows remarkable transferability and superior performance when handling unseen scenarios, demonstrating the strength of its design for multi-UAV navigation tasks in dynamic and complex environments.

\begin{table}[!h] 
\caption{Test metrics on performing navigation tasks in seen scenarios.\label{tab2}}
\begin{adjustwidth}{-\extralength}{0.5cm}
    \begin{tabularx}{\fulllength}{C|C|CCCCCC}
    \toprule
    \textbf{Metric}	 &\textbf{Method}	& Scene-I ($10\%$)	&Scene-I ($50\%$) & Scene-II ($10\%$)	& Scene-II ($50\%$) & Scene-III ($10\%$)	& Scene-III ($50\%$) \\
    \midrule
    \multirow{4}{*}{Avg. Transfer}  &MADDPG     &70.25  & 62.50   & 80.51  & 60.95   & 90.12	& 69.02  \\
    
   \multirow{4}{*}{Reward}  &MARDPG    &101.34  & 90.83   & 111.24  & 90.35   &97.18	&80.28 \\
    
   &MAPPO   &175.51  & 160.04   & 205.73  & 157.12   & 170.29	& 137.51 \\
    
    &\textbf{DTPPO}	& \textbf{262.89} & \textbf{251.77}    & \textbf{245.61}  &\textbf{235.19}   & \textbf{239.85}  & \textbf{221.49} \\

    \midrule
    \multirow{4}{*}{Avg. Collision} 
    & MADDPG     &4.95  & 23.71  & 7.69  & 22.11   &12.86	& 31.44\\
    \multirow{4}{*}{Penalty} & MARDPG    &3.35  & 15.18  & 7.73  & 18.53  &9.82	& 26.18 \\
    & MAPPO     & 2.41 & 4.28  &3.10  &5.31   & 4.65 & 7.12 \\
    
    &\textbf{DTPPO}	& \textbf{1.13} & \textbf{1.53}  & \textbf{1.12}  & \textbf{2.34}   & \textbf{4.21} &  \textbf{5.37} \\
    \midrule
    
    \multirow{4}{*}{Avg. Free} 
    &MADDPG     &1.42  & 1.06  & 1.95  &0.53  &0.72	& 0.40\\
    
    \multirow{4}{*}{Space Reward} &MARDPG   &1.31  & 1.63 & 1.94  & 1.10  &1.21	& 0.61 \\
    
    &MAPPO   & 3.76 & 2.98  &2.95 &\textbf{4.69}   & 2.07 & 1.90 \\
    
    &\textbf{DTPPO}	& \textbf{4.52} & \textbf{3.88}  & \textbf{4.96}  & 4.39   & \textbf{3.26} &  \textbf{3.11} \\
    
    \bottomrule
    \end{tabularx}
\end{adjustwidth}
\end{table}

\begin{figure}[H]
\begin{adjustwidth}{-\extralength}{0cm} % 根据需要调整外边距
    \centering
    \subfloat[Scene-I (10\%)]{
    \label{subfig:scene-I_0_1}
    \includegraphics[width=.31\linewidth]{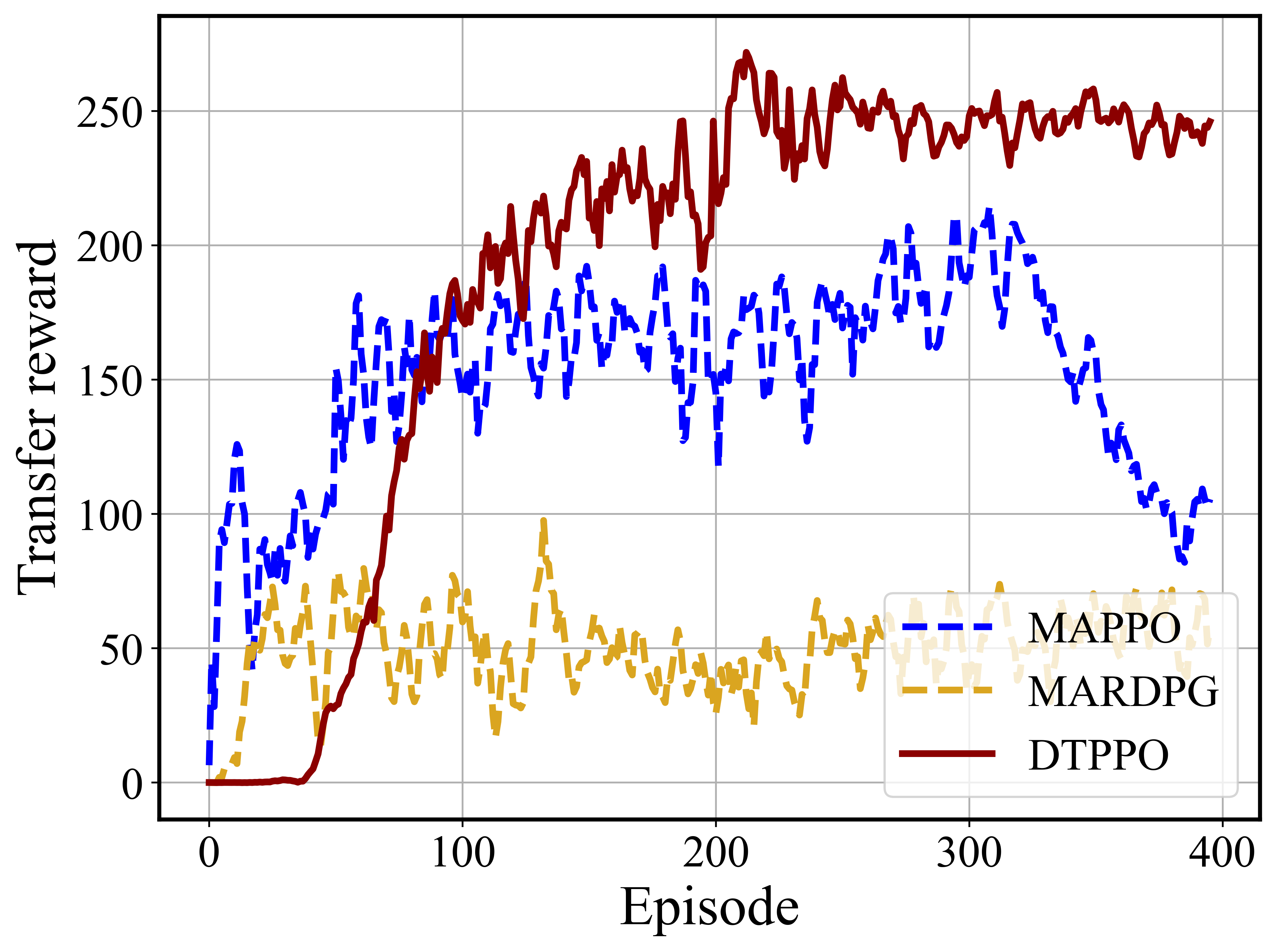}
    }\hfill
    % \hspace{0.15pt}
    \subfloat[Scene-II (10\%)]{
      \label{subfig:scene-II_0_1}
      \includegraphics[width=.31\linewidth]{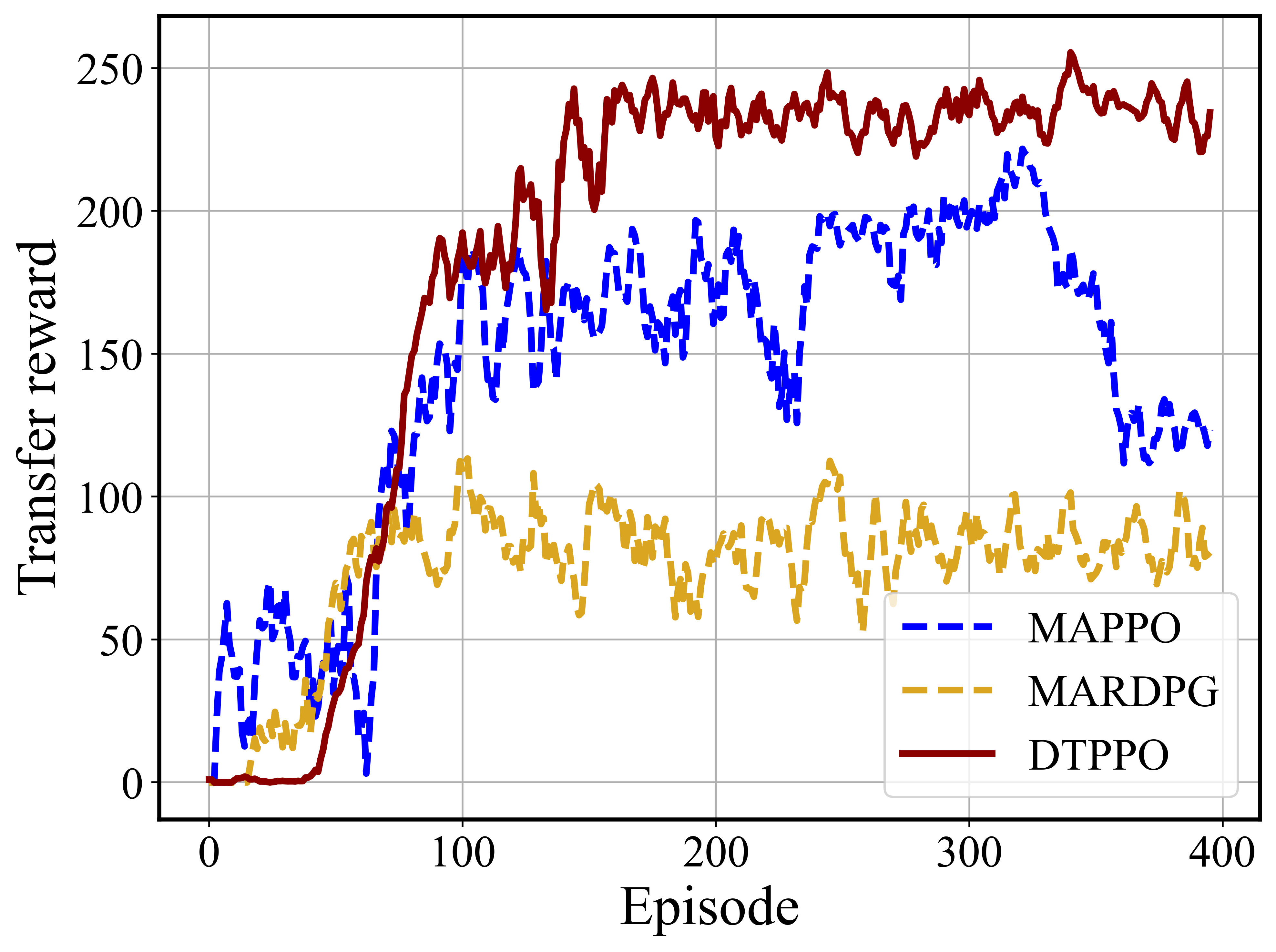}
      }\hfill
    % \hspace{0.15pt}
    \subfloat[Scene-III (10\%)]{
      \label{subfig:scene-III_0_1}
      \includegraphics[width=.31\linewidth]{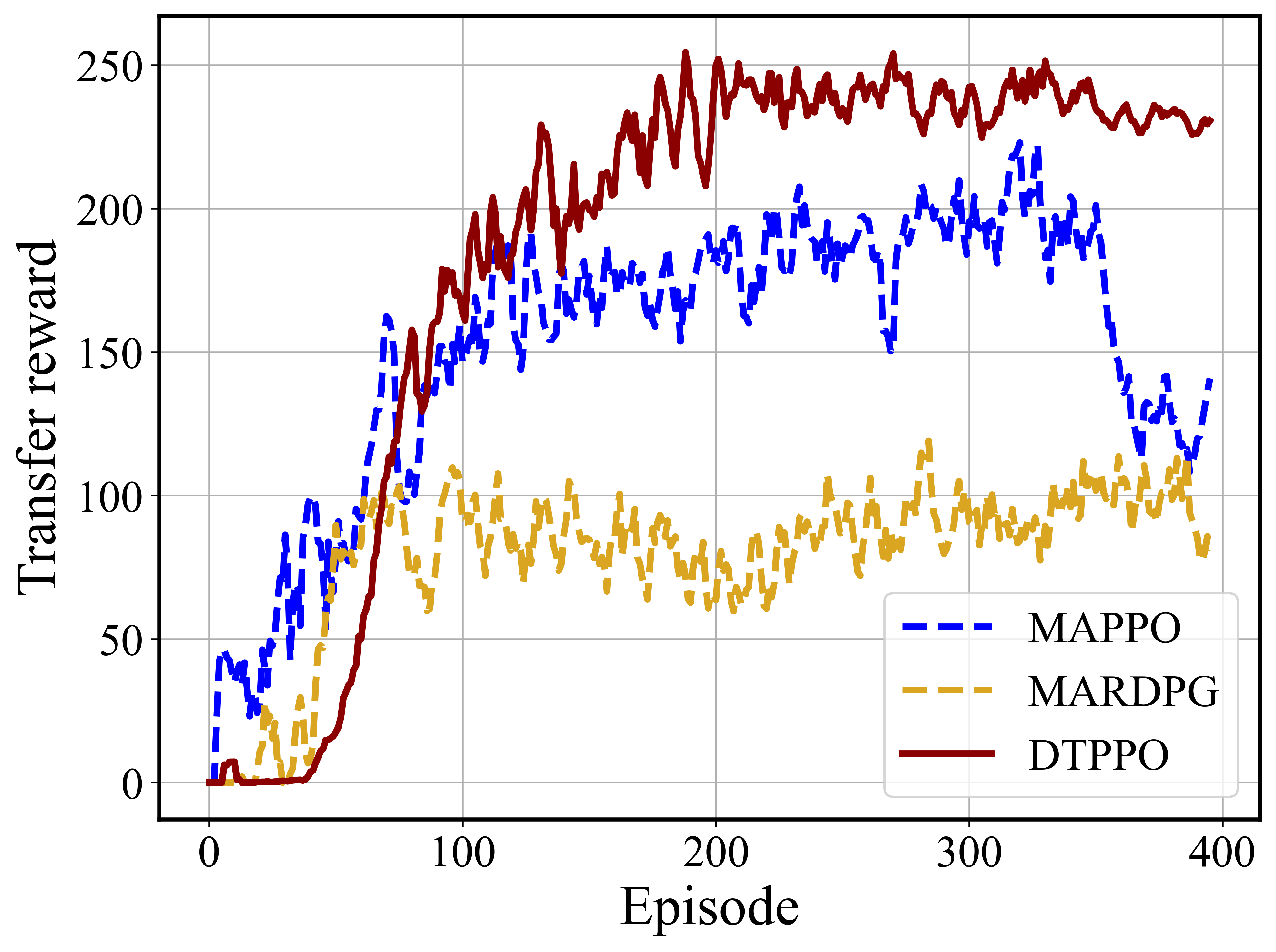}
      }
      
  \vspace{1pt}
    \subfloat[Scene-I (50\%)]{
      \label{subfig:scene-I_0_5}
      \includegraphics[width=.31\linewidth]{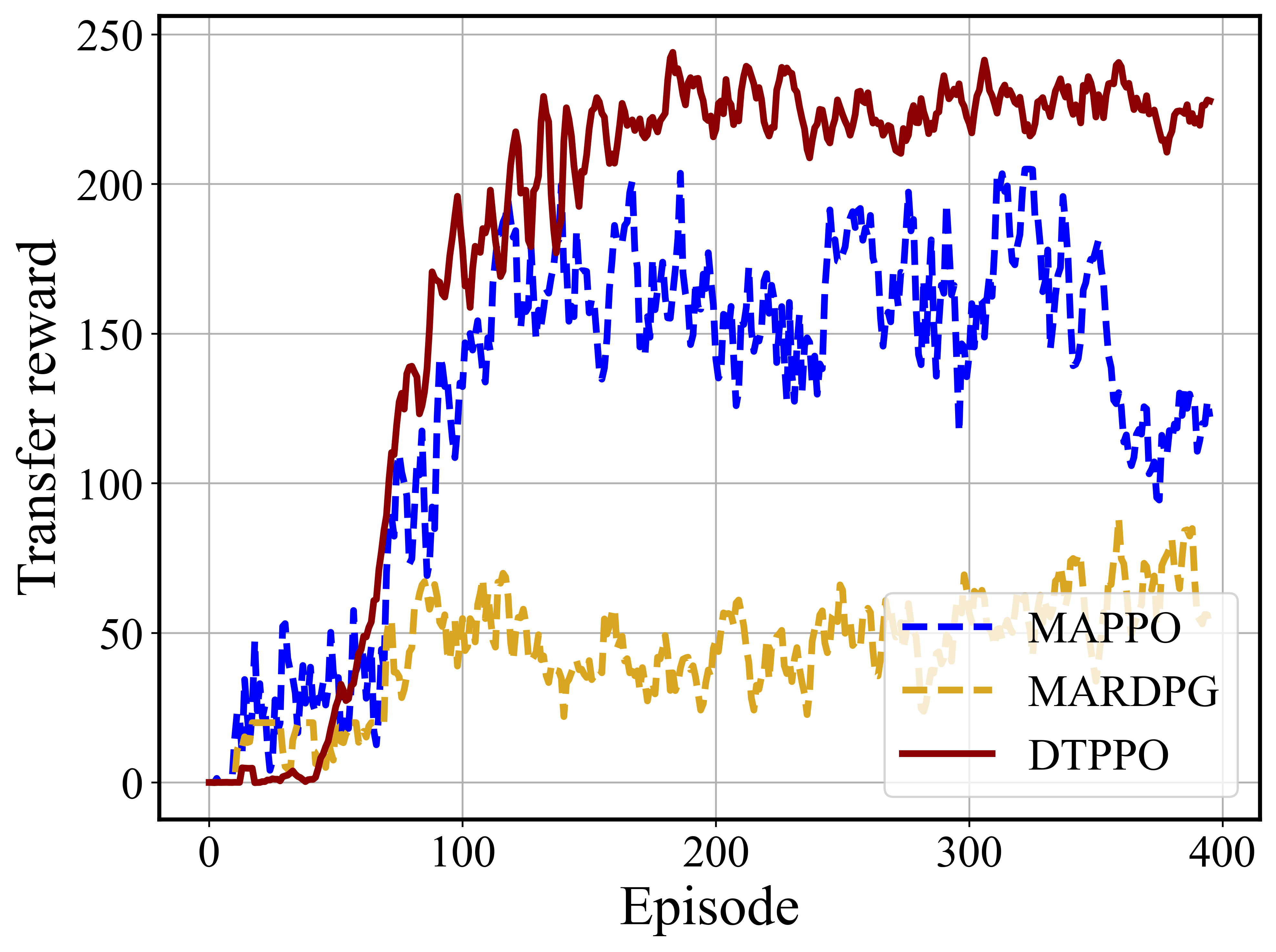}
      }\hfill
    % \hspace{0.15pt}
    \subfloat[Scene-II (50\%)]{
    \label{subfig:scene-II_0_5}
    \includegraphics[width=.31\linewidth]{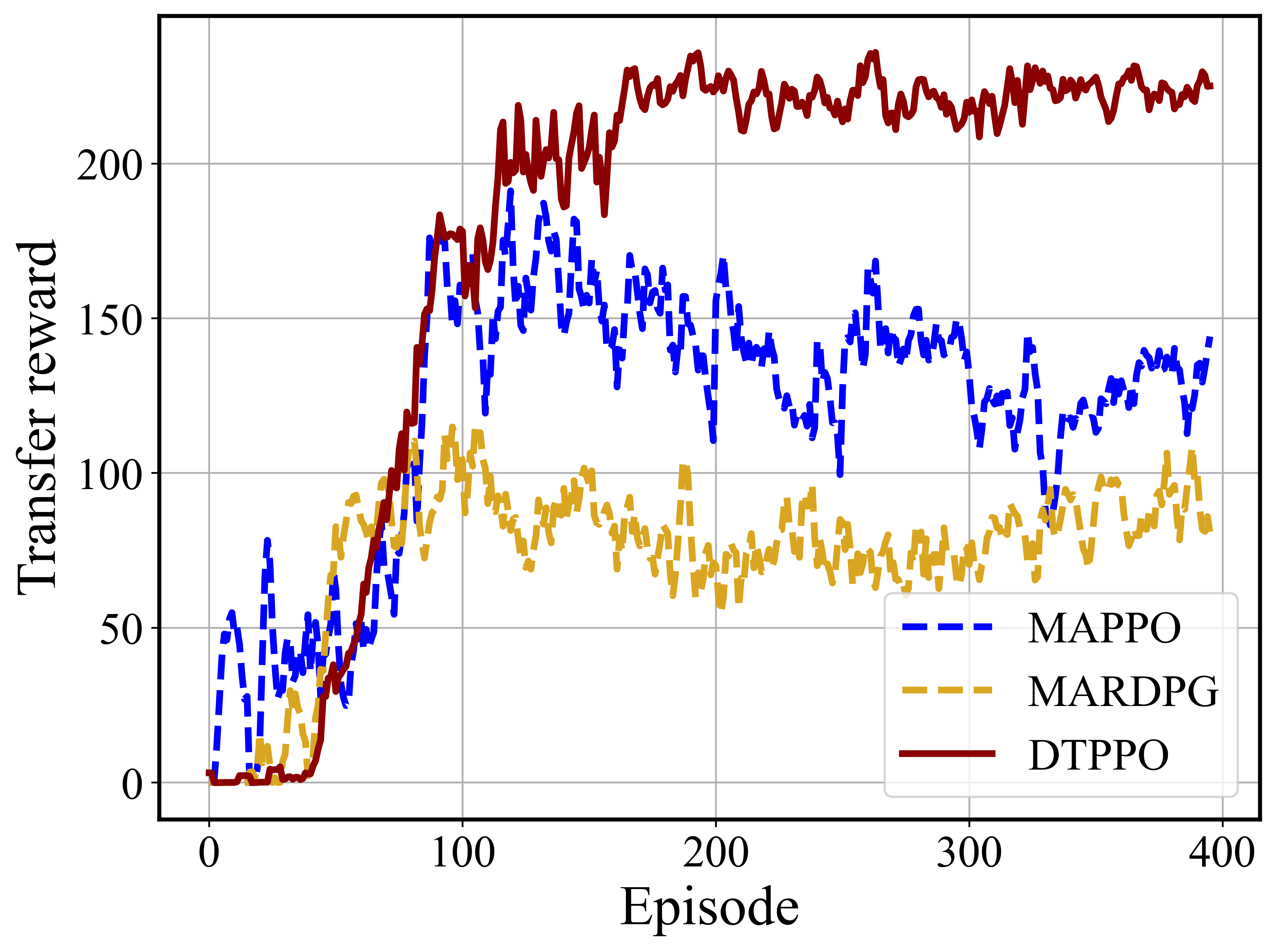}
    }\hfill
    % \hspace{0.25pt}
    \subfloat[Scene-III (50\%)]{
    \label{subfig:scene-III_0_5}
    \includegraphics[width=.31\linewidth]{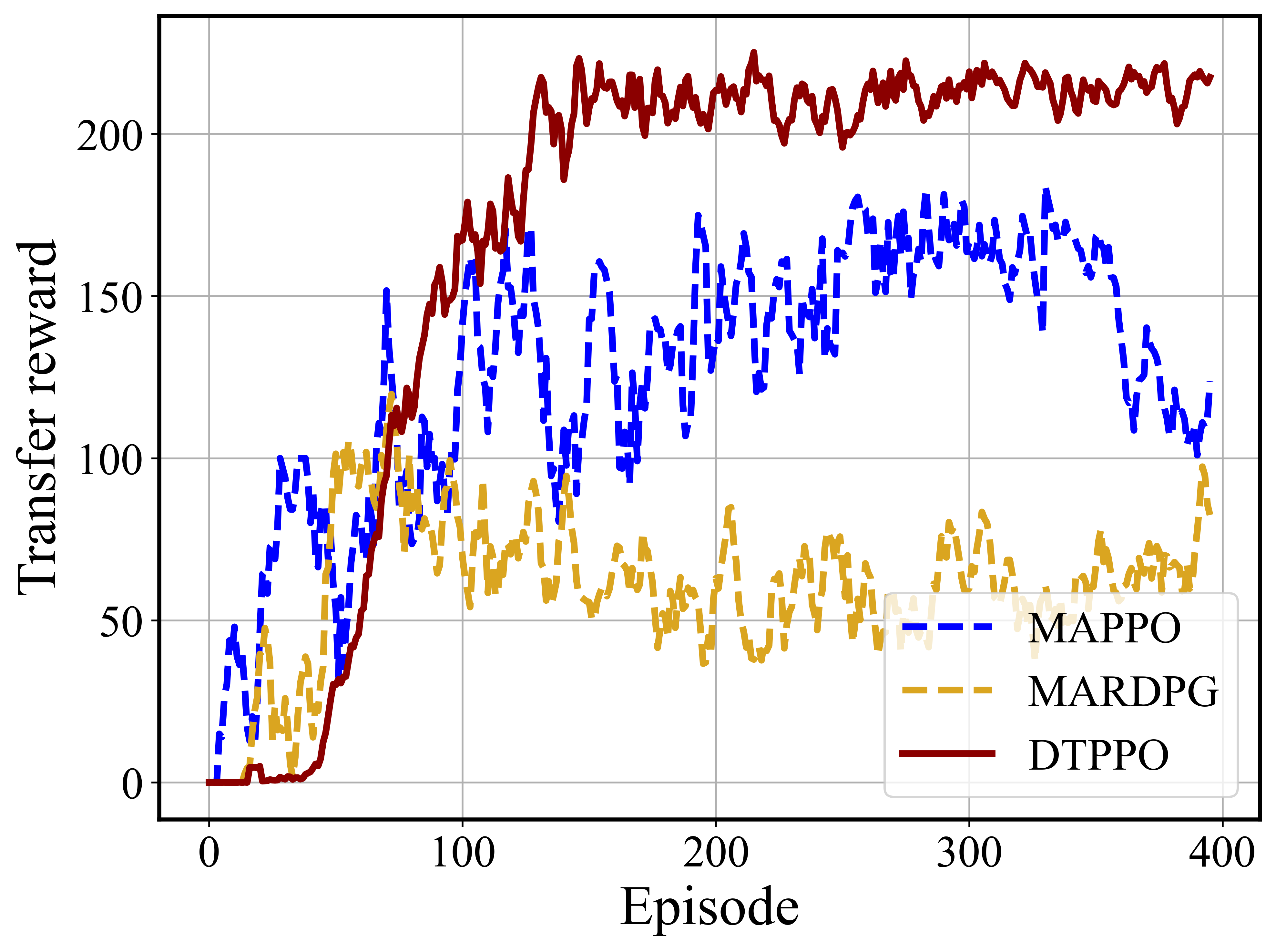}
    }
\end{adjustwidth}
\caption{Transfer reward during training.\label{exp_fig_1}}
\end{figure}

\subsubsection{Performance in Non-transfer Setting}
In this non-transfer setting, as shown in Table~\ref{tab2}, each scenario for testing is already seen during training. 
% Specifically, all methods are exposed to the three types of UAV environments in both 10\% and 50\% obstacle densities. Thus, this setup allows for evaluating the models' performance in familiar environments where generalization is not the primary concern but achieving the highest possible reward is.
Our method, still achieves the best results in all seen scenarios, demonstrating enhanced performance over MAPPO and MARDPG. 
For example, in the Scene-I (10\%) case, DTPPO yields an average transfer reward of 262.89, which is significantly higher than MAPPO's 175.51 and MARDPG's 101.34. This improvement is consistent across all other scenarios, showing DTPPO's robustness even in non-transfer settings. 
% On average, DTPPO exhibits a \textit{+6.20\%} gain in performance compared to MAPPO and a \textit{+9.96\%} improvement over MARDPG.
Moreover, the performance drop observed in Scene-II (50\%) and Scene-III (50\%) can be attributed to the higher complexity of these environments with denser obstacles. DTPPO consistently outperforms the other baselines by maintaining superior exploration capabilities, as reflected in its higher transfer rewards and free space rewards.
In terms of collision penalty, DTPPO registers the lowest penalty values across all scenarios, indicating safer navigation capabilities compared to MAPPO and MARDPG.

Furthermore, Figure~\ref{exp_fig_1} shows the transfer reward optimization process for the top 3 methods. 
DTPPO consistently outperforms the other two approaches in terms of both convergence speed and final performance.
The learning curves also highlight the stability of DTPPO during training, particularly in more challenging environments like Scene-II (50\%) and Scene-III (50\%), where MAPPO and MARDPG struggle with higher variance. In conclusion, DTPPO's ability to maintain high performance in both non-transfer and transfer settings, along with its superior learning stability, makes it an ideal solution for UAV navigation tasks in various obstacle-dense environments.

\begin{figure}[H]
% \begin{adjustwidth}{-\extralength}{0cm}
\centering
\includegraphics[width=1.0\linewidth]{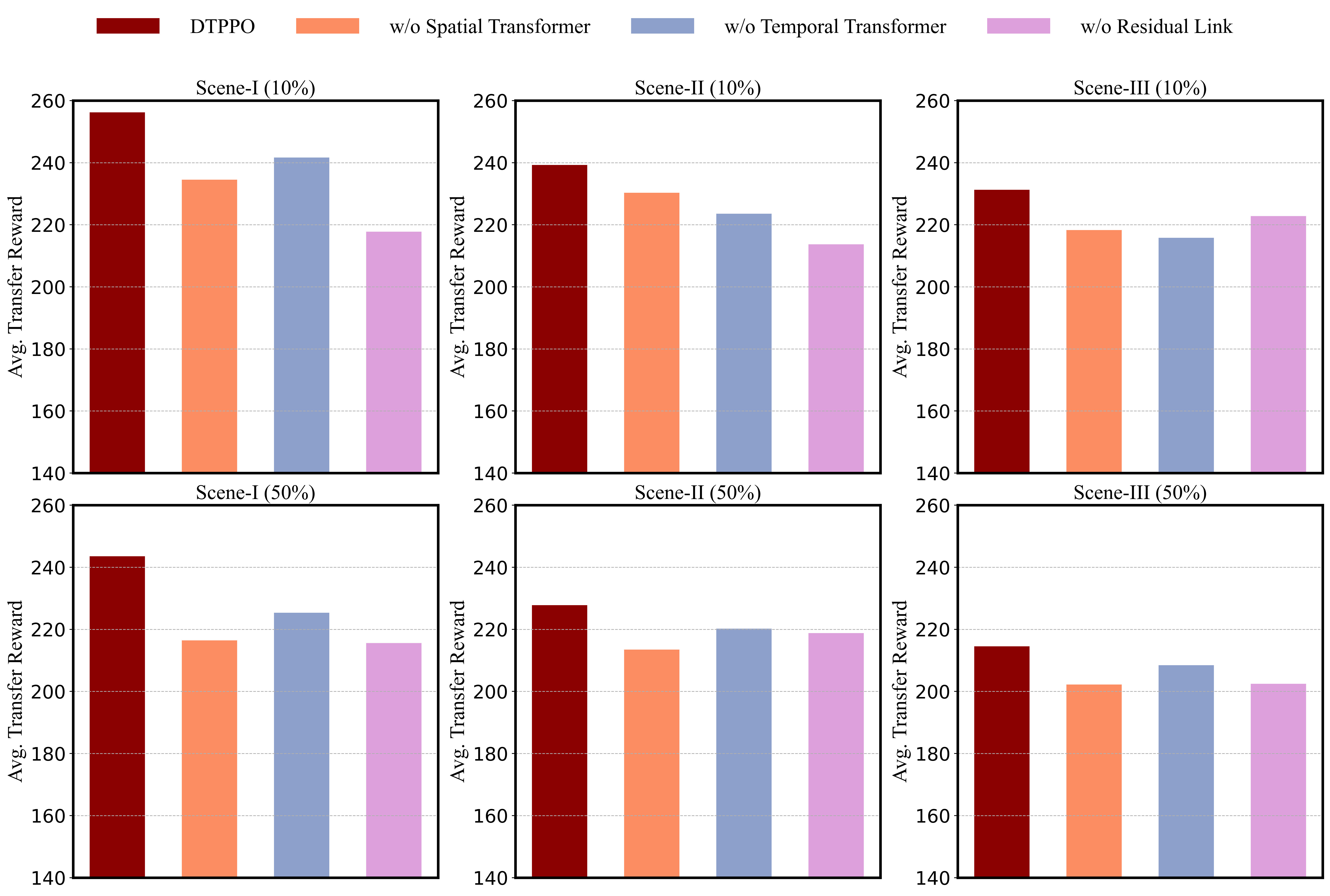}
% \end{adjustwidth}
\caption{Ablation study on different components in DTPPO.\label{ab}}
\end{figure}  

\subsubsection{Ablation Study}
In our ablation study, we investigate the impact of removing key components from the DTPPO framework. The results in Figure \ref{ab} illustrate the performance drop across six different test scenarios when excluding each of the following components:
\begin{itemize}
    \item \textit{w/o Spatial Transformer}: The removal of the spatial transformer, which facilitates inter-agent collaboration, results in the most significant drop in average transfer reward, especially in dense environments such as \textit{Scene-II (50\%)} and \textit{Scene-III (50\%)}. This emphasizes the critical role of spatial collaboration in complex, obstacle-filled environments.
    
    \item \textit{w/o Temporal Transformer}: Replacing the temporal transformer with a GRU leads to a noticeable decline in performance, particularly in scenarios like \textit{Scene-II (50\%)}. The ability to model temporal dependencies is crucial for maintaining high transfer rewards.

    \item \textit{w/o Residual Link}:  Removing the residual link significantly reduces performance across all scenarios, with the most pronounced drops observed in \textit{Scene-II (50\%)} and \textit{Scene-III (50\%)}. In these scenarios, the transfer reward decreases sharply compared to the full model, underscoring the critical role of self-observation in dense environments. Without the residual link, the model loses the ability to incorporate immediate feedback from its own state, resulting in less accurate decision-making and reduced performance, especially in more challenging environments.
    
\end{itemize}

\subsubsection{Varying Numbers of Scenarios}
\begin{figure}[H]
\begin{adjustwidth}{-\extralength}{0cm} % 根据需要调整外边距
    \centering
    \subfloat{
    \label{subfig:gain_Scene-I}
    \includegraphics[width=.31\linewidth]{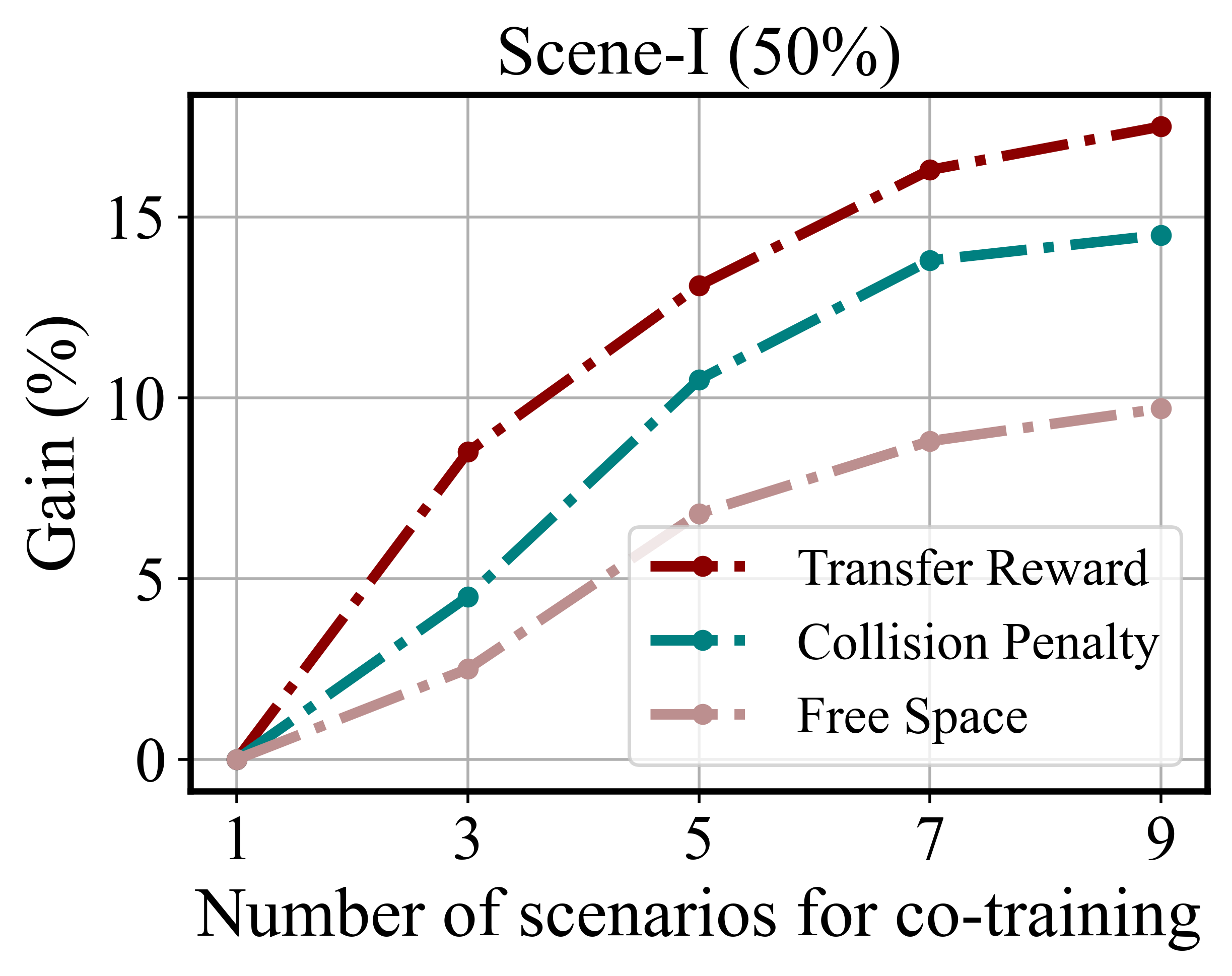}
    }\hfill
    % \hspace{0.15pt}
    \subfloat{
      \label{subfig:gain_Scene-II}
      \includegraphics[width=.31\linewidth]{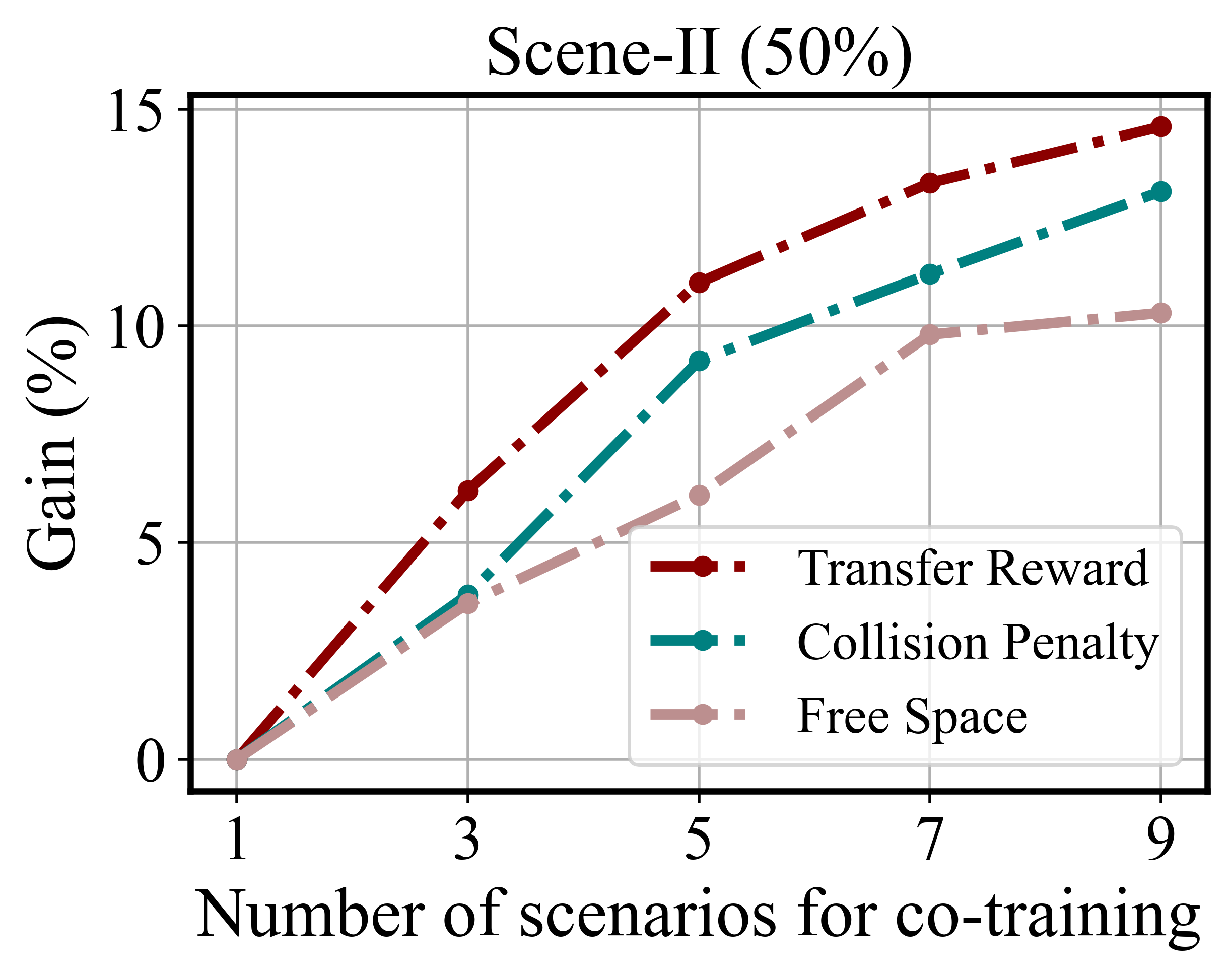}
      }\hfill
    % \hspace{0.15pt}
    \subfloat{
      \label{subfig:gain_Scene-III}
      \includegraphics[width=.31\linewidth]{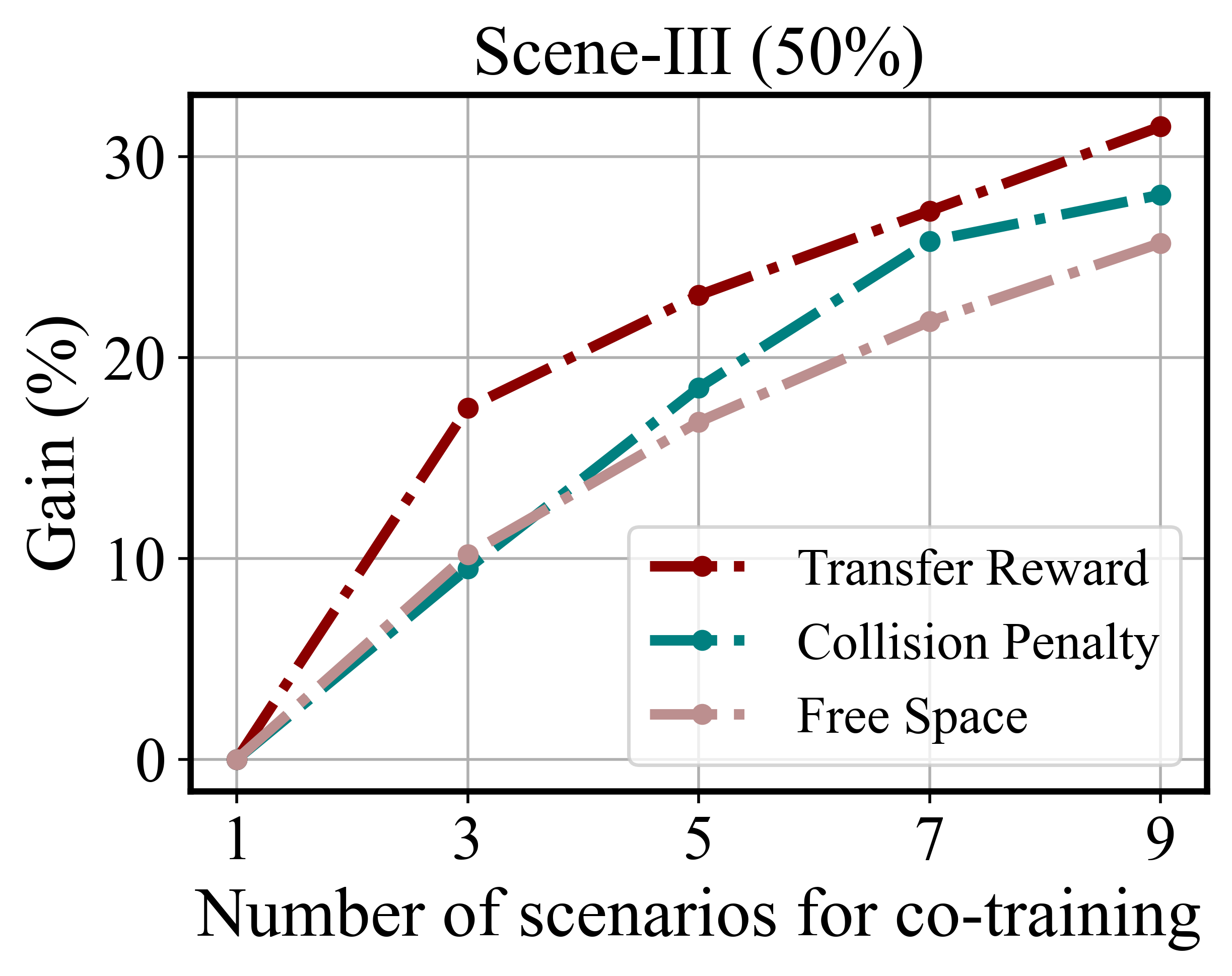}
      }
\end{adjustwidth}
\caption{Impact of varying the number of scenarios for co-training.\label{exp_fig_2}}
\end{figure}

We vary the number of scenarios for co-training from $[1, 3, 5, 7, 9]$ and investigate the impact on three unseen test scenarios with identical obstacle density: \textit{Scene-I (50\%)}, \textit{Scene-II (50\%)}, and \textit{Scene-III (50\%)}. The primary goal of this setting is to explore how increasing the diversity of co-training scenarios enhances our model’s ability to transfer effectively to dense environments. 
Figure \ref{exp_fig_2} shows the performance improvement on three test metrics. 
As the number of co-training scenarios increases, our model consistently achieves better performance. 
The gain in Transfer Reward grows steadily, reflecting improved adaptability to unseen dense environments. The Collision Penalty sees a significant reduction, indicating enhanced safety and collision avoidance capabilities. Although the Free Space Reward exhibits a more gradual increase, it still benefits from the larger set of co-training maps, further solidifying the overall robustness of our method in complex scenarios.

\begin{figure}[H]
    \centering
    \subfloat{
    \label{subfig:before train}
    \includegraphics[width=.475\linewidth]{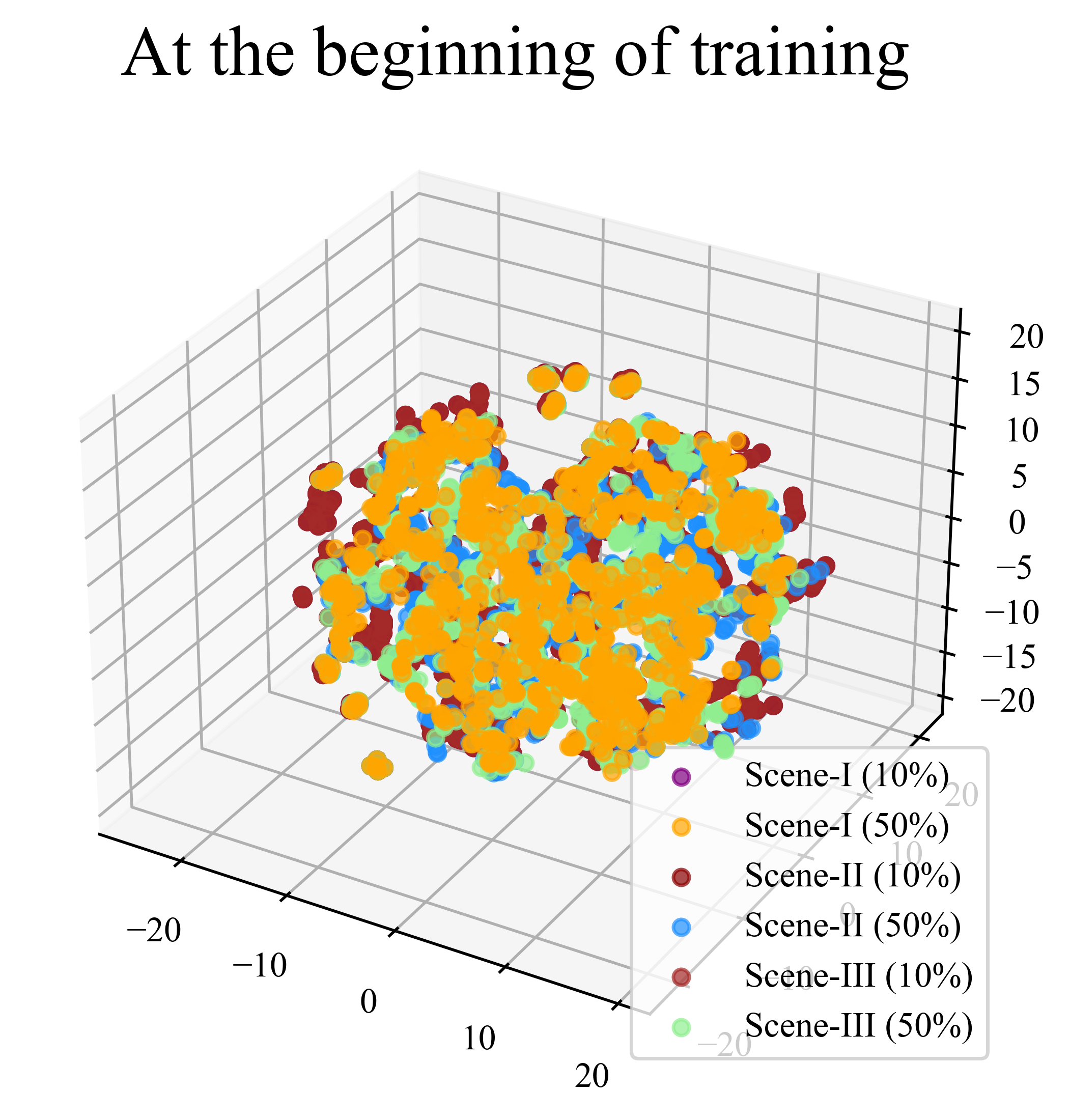}
    }\hfill
    % \hspace{0.15pt}
    \subfloat{
      \label{subfig:after train}
      \includegraphics[width=.475\linewidth]{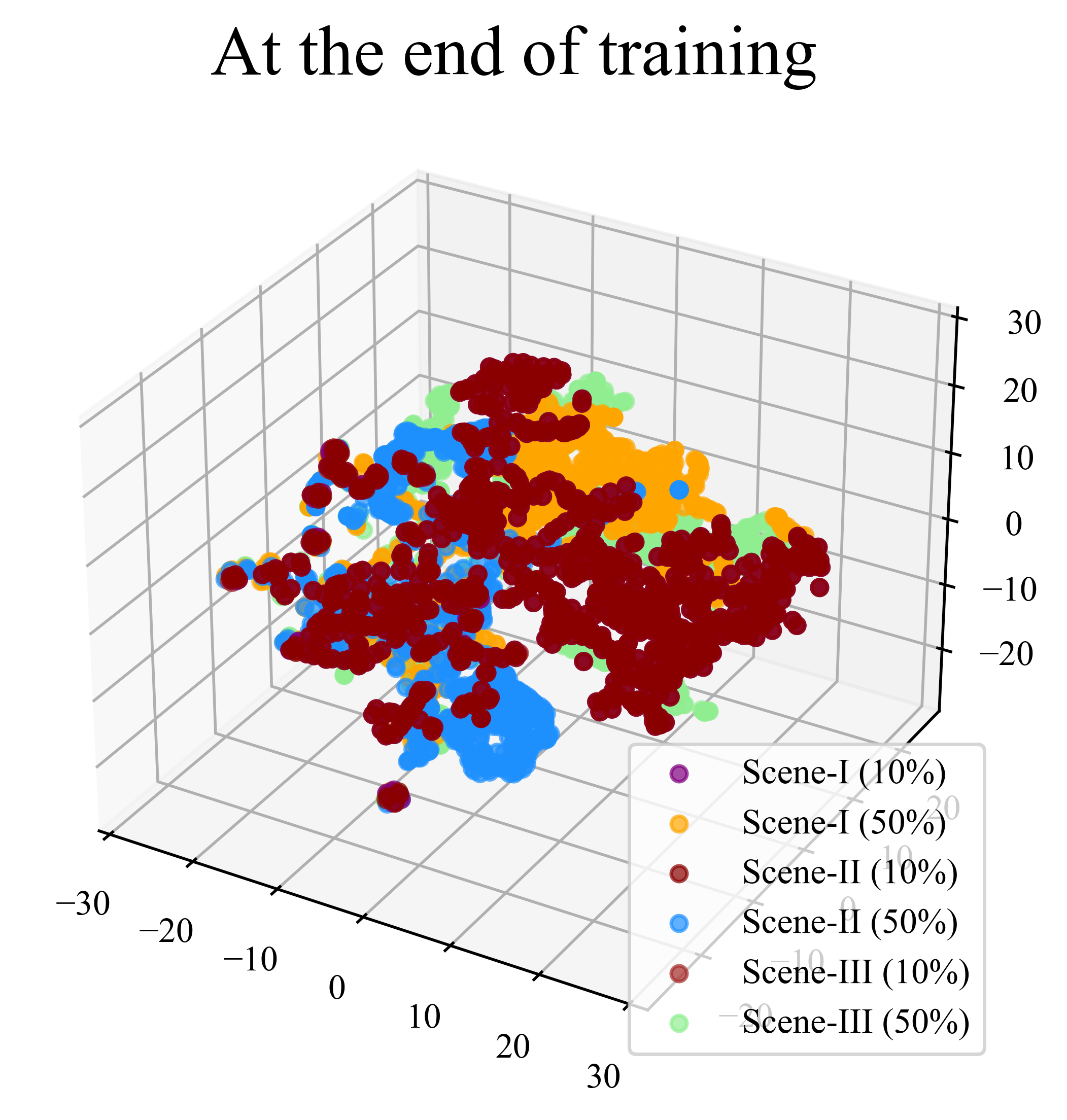}
      }\hfill
    % \hspace{0.15pt}
\caption{Visualizing Temporal Transformer’s output evaluated on Scene-III (50\%).\label{Visual_TT}}
\end{figure}
%%%%%%%%%%%%%%%%%%%%%%%%%%%%%%%%%%%%%%%%%%

\subsubsection{Analysis on Dual-T Encoder}
The Dual-T Encoder in DTPPO is a critical component that facilitates the model’s ability to capture both spatial and temporal dynamics in multi-agent environments.
We analyze the output embeddings from the Dual-T Encoder by visualizing the clustering patterns.
We apply the 3D t-SNE technique to visualize the clustering patterns. 
As shown in Figure \ref{Visual_TT}, When finishing training, the Dual-T Encoder is capable of grouping embeddings based on their respective scenarios, each represented by a distinct color. 
This result illustrates that the Dual-T Encoder can accurately capture scenario-specific dynamics information.

% We first visualize the attention weights of the Spatial Transformer, where we select 2 heads from 3 layers of the model on Scene-III (50\%).
% n Figure \ref{}, attention in Layer 1 is primarily focused on observations $o$ (green) around the target UAV and its neighbors.
% Moving to Layer 2, more attention is paid to actions $a$ (blue), suggesting that the model integrates higher-level information to guide action control for UAVS. 
% In Layer 3, all $o, a, r$ have high attention, especially the action $a$ and reward $r$, as the model prepares to output the final decision. 
% This progressive adjustment of attention from low-level observations to high-level rewards and actions illustrates the effectiveness of the Spatial Transformer in modeling intricate interrelations between $o,a,r$.

% We also analyze the output embeddings from the Temporal Transformer by visualizing the clustering patterns.
% We apply the 3D t-SNE technique to visualize the clustering patterns. 
% As shown in Figure \ref{Visual_TT}, after training, the Temporal Transformer is capable of grouping embeddings based on their respective scenarios, each represented by a distinct color. 
% This result illustrates that the Temporal Transformer can accurately capture scenario-specific dynamics information.

\section{Conclusions}
\label{Conclusions}
In this paper, we proposed DTPPO, a Dual-Transformer Encoder-based PPO method aimed at solving the challenge of multi-UAV navigation in unseen complex environments. 
By integrating a Spatial Transformer to enhance inter-UAV coordination and a Temporal Transformer to model temporal dynamics, DTPPO improves both navigation efficiency and transferability. 
Our experimental results across various obstacle-laden environments validate the superior performance of DTPPO over baseline methods, particularly in unseen scenarios where the system demonstrates robust transfer capabilities. 
Notably, the framework significantly reduces the need for scenario-specific retraining, minimizing computational costs and enabling real-time adaptability.
Future work will focus on further enhancing transfer learning techniques to address increasingly dynamic environments and real-world deployment scenarios with more heterogeneous UAV fleets.

%%%%%%%%%%%%%%%%%%%%%%%%%%%%%%%%%%%%%%%%%%
% \section{Patents}

% This section is not mandatory, but may be added if there are patents resulting from the work reported in this manuscript.

%%%%%%%%%%%%%%%%%%%%%%%%%%%%%%%%%%%%%%%%%%
\vspace{6pt} 

%%%%%%%%%%%%%%%%%%%%%%%%%%%%%%%%%%%%%%%%%%
%% optional
%\supplementary{The following supporting information can be downloaded at:  \linksupplementary{s1}, Figure S1: title; Table S1: title; Video S1: title.}


%\begin{tabularx}{\textwidth}{lll}
%\toprule
%\textbf{Name} & \textbf{Type} & \textbf{Description} \\
%\midrule
%S1 & Python script (.py) & Script of python source code used in XX \\
%S2 & Text (.txt) & Script of modelling code used to make Figure X \\
%S3 & Text (.txt) & Raw data from experiment X \\
%S4 & Video (.mp4) & Video demonstrating the hardware in use \\
%\bottomrule
%\end{tabularx}
%}

%%%%%%%%%%%%%%%%%%%%%%%%%%%%%%%%%%%%%%%%%%
\authorcontributions{Conceptualization, Jintao Liang, Anning Wei, Ziyue Li, Rui Zhao; 
methodology, Jintao Liang, Anning Wei, Ziyue Li; 
software, Jintao Liang, Anning Wei; 
validation, Jintao Liang, Anning Wei; 
formal analysis, Jintao Liang, Anning Wei; 
investigation, Jintao Liang, Anning Wei; 
resources, Ziyue Li, Rui Zhao; 
data curation, Jintao Liang; 
writing---original draft preparation, Jintao Liang, Anning Wei; 
writing---review and editing, Ziyue Li, Kaiyuan Lin; 
visualization, Jintao Liang, Anning Wei; 
supervision, Ziyue Li; 
project administration, Ziyue Li, Rui Zhao.
% funding acquisition, Y.Y. 
All authors have read and agreed to the published version of the manuscript.}

\funding{This research received no external funding.}

\conflictsofinterest{
The authors declare no conflicts of interest.
}
%%%%%%%%%%%%%%%%%%%%%%%%%%%%%%%%%%%%%%%%%%
%% Optional

%% Only for journal Encyclopedia
%\entrylink{The Link to this entry published on the encyclopedia platform.}

% \abbreviations{Abbreviations}{
% The following abbreviations are used in this manuscript:\\

% \noindent 
% \begin{tabular}{@{}ll}
% MDPI & Multidisciplinary Digital Publishing Institute\\
% DOAJ & Directory of open access journals\\
% TLA & Three letter acronym\\
% LD & Linear dichroism
% \end{tabular}
% }

%%%%%%%%%%%%%%%%%%%%%%%%%%%%%%%%%%%%%%%%%%
%% Optional
\appendixtitles{no} % Leave argument "no" if all appendix headings stay EMPTY (then no dot is printed after "Appendix A"). If the appendix sections contain a heading then change the argument to "yes".
% \appendixstart
% \appendix
% \section[\appendixname~\thesection]{}
% \subsection[\appendixname~\thesubsection]{}
% The appendix is an optional section that can contain details and data supplemental to the main text---for example, explanations of experimental details that would disrupt the flow of the main text but nonetheless remain crucial to understanding and reproducing the research shown; figures of replicates for experiments of which representative data are shown in the main text can be added here if brief, or as Supplementary Data. Mathematical proofs of results not central to the paper can be added as an appendix.

% \begin{table}[H] 
% \caption{This is a table caption.\label{tab5}}
% \newcolumntype{C}{>{\centering\arraybackslash}X}
% \begin{tabularx}{\textwidth}{CCC}
% \toprule
% \textbf{Title 1}	& \textbf{Title 2}	& \textbf{Title 3}\\
% \midrule
% Entry 1		& Data			& Data\\
% Entry 2		& Data			& Data\\
% \bottomrule
% \end{tabularx}
% \end{table}

% \section[\appendixname~\thesection]{}
% All appendix sections must be cited in the main text. In the appendices, Figures, Tables, etc. should be labeled, starting with ``A''---e.g., Figure A1, Figure A2, etc.

%%%%%%%%%%%%%%%%%%%%%%%%%%%%%%%%%%%%%%%%%%
\begin{adjustwidth}{-\extralength}{0cm}

\reftitle{References}

\PublishersNote{}
\end{adjustwidth}
\end{document}